\documentclass[12pt]{article}
\usepackage{bbm}
\usepackage{color}
\usepackage{authblk}
\usepackage{latexsym}
\usepackage{epsfig,amssymb,euscript, mathrsfs,cite}
\usepackage{amsmath}
\usepackage{slashed}
\usepackage{mathrsfs} 
\usepackage{enumerate}
\definecolor{MyDarkBlue}{rgb}{0.15,0.15,0.45}
\usepackage[linktocpage=true]{hyperref}
\hypersetup{
colorlinks=true,
citecolor=MyDarkBlue,
linkcolor=MyDarkBlue,
urlcolor=MyDarkBlue,
pdfauthor={},
pdftitle={},
pdfsubject={hep-th}
}
\newsavebox{\ns}
\newsavebox{\dbrane}
\newsavebox{\dbshort}

\def\be{\begin{equation}}
\def\ee{\end{equation}}
\def\bea{\begin{eqnarray}}
\def\eea{\end{eqnarray}}

\newcommand{\nn}{\nonumber\\}

\newcommand\Z{\mathbb{Z}}

\newcommand\C{\mathbb{C}}

\newcommand\diff{\mathrm{d}}

\newcommand{\ii}{\mathrm{i}}

\newcommand{\ex}{\mathrm{e}}

\newcommand{\z}{z}

\newlength{\sswidth}

\numberwithin{equation}{section}      

\newcommand{\jp}{\mathtt{j}}

\newcommand{\deltap}{\delta}
\newcommand{\sigmap}{g^2}
\newcommand{\xp}{x}

\begin{document}

\begin{titlepage}

\vskip 1.5cm

\begin{center}


{\Large \bf Multi-charge accelerating black holes\\[3mm] and  spinning spindles}

\vskip 1cm

\vskip 1cm
{Pietro Ferrero$^{\mathrm{a}}$,  Matteo Inglese$^{\mathrm{b}}$,  Dario Martelli$^{\mathrm{b,c,d}}$, and James Sparks$^{\mathrm{a}}$}

\vskip 0.5cm

${}^{\,\mathrm{a}}$\textit{Mathematical Institute, University of Oxford,\\
Andrew Wiles Building, Radcliffe Observatory Quarter,\\
Woodstock Road, Oxford, OX2 6GG, U.K.\\}

\vskip 0.2cm

${}^{\mathrm{b}}$\textit{Dipartimento di Matematica ``Giuseppe Peano'', Universit\`a di Torino,\\
Via Carlo Alberto 10, 10123 Torino, Italy}

\vskip 0.2cm

${}^{\mathrm{c}}$\textit{INFN, Sezione di Torino \&}   ${}^{\mathrm{d}}$\textit{Arnold--Regge Center,\\
 Via Pietro Giuria 1, 10125 Torino, Italy}

\end{center}

\vskip 2 cm

\begin{abstract}
\noindent  We construct a family of multi-dyonically charged and rotating supersymmetric AdS$_2\times \Sigma$ solutions of $D=4$, $\mathcal{N}=4$ gauged supergravity, where $\Sigma$ is a sphere with two conical singularities known as a spindle. We argue 
that these arise as near horizon limits of extremal dyonically charged rotating and accelerating supersymmetric black holes in AdS$_4$, that we conjecture to exist.  We demonstrate this in the non-rotating limit, 
 constructing the accelerating black hole solutions and showing that the non-spinning spindle solutions arise as the near horizon limit of the supersymmetric and extremal sub-class of these black holes.
From the near horizon solutions we compute the Bekenstein-Hawking entropy of the black holes as a function
of the conserved charges, and  show that this may equivalently be obtained by extremizing a simple entropy 
function.  For appropriately quantized  magnetic fluxes, the solutions uplift on $S^7$, or its ${\cal N}=4$ orbifolds $S^7/\Gamma$, to smooth supersymmetric solutions to $D=11$ supergravity, where
the entropy  is expected to count microstates of the theory on $N$ M2-branes wrapped on a spinning 
spindle, in the large $N$ limit.

\end{abstract}

\end{titlepage}

\pagestyle{plain}
\setcounter{page}{1}
\newcounter{bean}
\baselineskip18pt

\tableofcontents

\newpage

\section{Introduction}\label{sec:intro}

Thanks to the AdS/CFT correspondence, in recent years there has been tremendous progress in elucidating the microscopic degrees of freedom for large classes of black holes.  In the context of $D=4$ dimensions, which is the focus of the present paper,  starting with  \cite{Benini:2015eyy,Benini:2016rke} a successful strategy for reproducing the Bekenstein-Hawking entropy of supersymmetric 
asymptotically locally AdS$_4$ black holes has been developed. In particular, the entropy can be extracted by analysing appropriate supersymmetric statistical ensembles of the dual $d=3$ SCFT, which are exactly calculable using localization
 techniques.  Two main classes of black holes have been discussed using this approach. One class consists of static black holes with hyperbolic horizons, for which the dual field theory is typically defined on $S^1\times \Sigma_g$, where $\Sigma_g$ is a genus $g$ Riemann surface equipped with a constant curvature metric, and in order to preserve supersymmetry one performs the so-called topological twist \cite{Benini:2015noa}. A second class consists of rotating Kerr-Newman-AdS black holes with spherical horizons, for which the dual field theory is defined on a ``spinning'' $S^1\times S^2$ \cite{Nian:2019pxj}.  

In \cite{Ferrero:2020twa} a different class of asymptotically locally AdS$_4$ black holes  has been considered in the context of holography. 
These are a family of solutions to Einstein-Maxwell theory with a cosmological constant, or equivalently minimal $D=4$, $\mathcal{N}=2$ gauged supergravity, 
originally constructed by Pleba\'nski and Demia\'nski \cite{Plebanski:1976gy,Podolsky:2006px}. 
The Pleba\'nski-Demia\'nski  solutions describe the most general dyonic, rotating and accelerating 
black holes in minimal gauged supergravity, and have a number of striking features. The term ``accelerating'' refers to the fact that the black hole curvature singularity can be shown 
to have a uniform proper acceleration, and more generally in a natural frame any 
world-line with constant space-like coordinates also has this property -- see, {\it e.g.} section III of \cite{Podolsky:2006px}. It is well-known that the acceleration is associated with conical deficit angles, which may be interpreted as being sourced by strings in the black hole geometry
\cite{PhysRevD.2.1359}\footnote{It is also well-known that such accelerating black holes emit gravitational radiation \cite{Podolsky:2003gm}, but the resulting energy loss is balanced by the force exerted by the strings, which keeps the acceleration constant.}. In 
\cite{Ferrero:2020twa} it has been shown that by embedding the black holes in $D=11$ supergravity, the conical singularities may be completely removed. From the four-dimensional point of view, the conical deficits manifest themselves as orbifold singularities on the horizon, which becomes a ``spindle''~$\Sigma$.

 An important property of these Pleba\'nski-Demia\'nski   solutions  is that, in the context of minimal gauged supergravity, they admit a supersymmetric and extremal sub-family of dyonic accelerating and rotating black holes \cite{Klemm:2013eca}, whose near horizon geometry is  a spinning  AdS$_2\times \Sigma$ solution \cite{Ferrero:2020twa}. There are two interesting limits of this family: (i)
turning off the acceleration parameter, one finds that the magnetic charge is also necessarily zero, and
the spindle $\Sigma$ becomes a two-sphere $S^2$. These are then the rotating Kerr-Newman-AdS black holes with spherical horizons 
mentioned above, and studied recently in \cite{Bobev:2019zmz, Nian:2019pxj}. (ii) On the other hand, instead turning off the rotation parameter, 
one finds the electric charge is now necessarily zero. The near horizon limits give  non-rotating AdS$_2\times \Sigma$ solutions, 
whose uplift to $D=11$ supergravity are AdS$_2$ solutions first constructed in \cite{Gauntlett:2006ns} using an entirely different approach. 
 Another remarkable aspect of the general family of dyonic accelerating and rotating solutions is that supersymmetry is realized in a novel way, which is distinct from the topological twist, even in the case of vanishing rotation \cite{Ferrero:2020laf}.  Interestingly,  the Bekenstein-Hawking entropy of the black holes as a function of the physical charges, as well as the spindle deficit angles, can be obtained purely from the near horizon solutions. 
The black hole solutions have been further studied in  \cite{Cassani:2021dwa}, where it has been shown that the entropy can be derived from a Legendre transform of the Euclidean on-shell action, similarly to 
 \cite{Cabo-Bizet:2018ehj,Cassani:2019mms}, thus setting the stage for a direct analysis of the dual $d=3$ SCFT, defined on   $S^1\times \Sigma$ (spinning or otherwise). 

In this paper we will discuss extensions of the above solutions to non-minimal supergravities, focusing on an ${\cal N}=4$ supergravity model that  arises as a consistent truncation of $D=11$ supergravity. Alternatively, this can be regarded 
as minimal ${\cal N}=2$ gauged supergravity coupled to one vector multiplet with a particular prepotential.  As such, solutions of this model can be uplifted to solutions of $D=11$ supergravity, and therefore interpreted holographically as dual to 
${\cal N}=4$ \cite{Gaiotto:2008ak} (or ${\cal N}=2$),  $d=3$ SCFTs arising on M2-branes.  We will also make some comments on solutions to  a more general model, known as  the STU model, containing 
two additional vector multiplets, which provides a general setting for solutions dual to the ABJM field theory 
on $N$ M2-branes.  

  Several solutions to this supergravity theory (or indeed to the STU model)  are known in the literature, and we have summarized those relevant for holography in the diagram in Figure~\ref{fig:diagram} below.  
\begin{figure}[hbt!]
\begin{center}
        \includegraphics[width=1\textwidth]{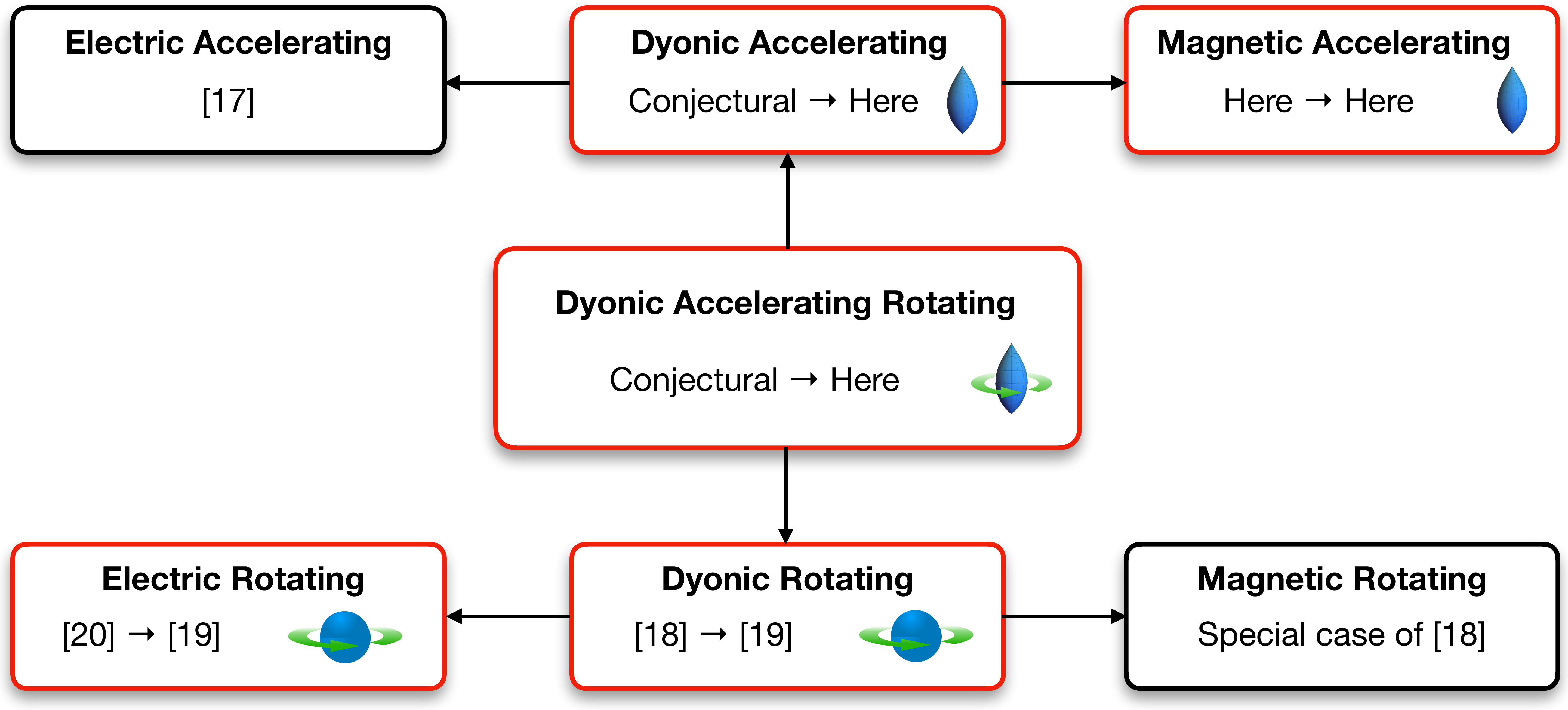}
\caption{Summary of AdS$_4$ black holes with either spherical or spindle horizons in $D=4$, ${\cal N}=4$ gauged supergravity.  The solutions in the red frames admit a supersymmetric and extremal limit and their near horizon AdS$_2\times \Sigma$ geometries are represented pictorially. From bottom-left to top-right: a spinning sphere, a spinning spindle, and a non-spinning spindle.  In all cases the reference on the left refers to  the non-extremal black holes, and that on the right refers to the near horizon solution in the supersymmetric limit. 
The spinning spindles (in the central box) admit two special limits: going up, rotation can be switched off; going down, acceleration can be switched off (it is also possible to switch off both acceleration and rotation, yielding a non-spinning sphere).
 The multi-charge spinning spindles are constructed in section \ref{sec:AdS2solutions}, while the magnetic accelerating black holes are constructed in section \ref{sec:BH}.}\label{fig:diagram}
\end{center}
\end{figure} 
In particular, there exist two notable classes of solutions. 
A  solution describing electrically charged, non-rotating, accelerating black holes was presented in \cite{Lu:2014sza}. This is a multi-charge 
generalization of the AdS $C$-metric in the Einstein-Maxwell theory, which is a member of the solutions in \cite{Plebanski:1976gy}, and  does not admit a supersymmetric limit.  Reference \cite{Chow:2013gba}  
constructed a dyonic, rotating, but non-accelerating, family of black hole solutions. Imposing supersymmetry on this family leads to a
dyonic rotating solution, recovered in \cite{Hristov:2019mqp}, that
also discusses the AdS$_2\times S^2$ near horizon solution.  In turn, switching off the magnetic charge, the solution reduces to a multi-electric charge
 version of the  Kerr-Newman black hole of Maxwell-Einstein theory, that was 
previously discovered in \cite{Chong:2004na}. The supersymmetric limit of this was  discussed in \cite{Cvetic:2005zi}, and is the multi-electric charge counterpart of the extremal 
Kerr-Newman black hole of Maxwell-Einstein theory \cite{Caldarelli:1998hg}.  
Based on these and on the family of black hole solutions of the minimal theory \cite{Plebanski:1976gy,Podolsky:2006px}, we conjecture that there should exist a family of 
multi-charge, dyonic, accelerating and rotating AdS$_4$ black holes, from which all the other solutions should arise as special cases. Unfortunately, we have not been able to construct this family, which remains a challenge for future work. However, we have constructed a number of new solutions, whose existence adds weight to this conjecture.

Firstly, we have constructed a family of supersymmetric multi-charge spinning spindle solutions, namely rotating AdS$_2\times \Sigma$ solutions, from which we computed the associated entropy, as a function of the physical charges and the spindle data.\footnote{A supersymmetric spinning spindle solution in the so-called $t^3$ model, has been presented in \cite{Faedo:2021kur}. However, this is not expected to arise as the near horizon limit of an extremal 
 AdS$_4$ black hole.} We expect this family to arise as the near horizon limit of the corresponding family of supersymmetric and extremal black holes discussed above. Secondly, when the rotation is turned off, we have constructed an explicit family of  magnetically charged, accelerating, but non-rotating AdS$_4$ black holes from which the non-spinning spindle solutions arise in the near horizon limit, after imposing supersymmetry and extremality.  In  the diagram in Figure~\ref{fig:diagram} we have summarized the relevant  previously known solutions, as well as the new solutions that we have constructed in this paper.
 Finally, we have proposed  an entropy function, from which the Bekenstein-Hawking entropy can be reproduced via a Legendre transform and imposing reality conditions on the charges, generalizing and unifying 
the entropy functions proposed in \cite{Cassani:2019mms},  \cite{Hristov:2019mqp}, \cite{Hosseini:2020mut}, and    \cite{Cassani:2021dwa}. 

The rest of the paper is organized as follows.  In section \ref{sec:AdS2solutions} we introduce the $D=4$ supergravity model that we shall consider,  and we construct new supersymmetric AdS$_2\times \Sigma$ solutions using a suitable ansatz.  We conjecture these to arise as the near horizon limits of rotating and accelerating black holes with two independent pairs of dyonic charges.  We analyse and solve the conditions that are required to have a smooth orbifold metric on a spindle $\Sigma=\mathbb{WCP}^1_{[n_-,n_+]}$,  and compute the conserved charges and entropy from the near horizon solution.  We also compute explicitly the Killing spinor,  and show that the solution is regular when uplifted to $D=11$ supergravity.  In section~\ref{sec:BH} we show that,  in the case with only magnetic charges,  the solutions of section~\ref{sec:AdS2solutions} are in fact the near horizon limits of supersymmetric,  accelerating,  magnetically charged black holes.  In section \ref{sec:conjectures} we make an educated guess for the on-shell action of the black holes conjectured in section \ref{sec:AdS2solutions},  which allows us to derive their entropy by extremizing a suitably defined entropy function. We also conjecture an expression  for  the mass of such black holes,  in the supersymmetric case.  Section \ref{sec:discussion} concludes with some open problems and possible extensions of our work. 

We also include a number of appendices.  In appendix \ref{app:ansatz} we give further details of the ansatz that leads to the solutions discussed in section \ref{sec:AdS2solutions},  as well as discussing some limiting cases of the general solution.  In appendix \ref{app:NHlimit} we show explicitly that the near horizon limit of the magnetic, accelerating black holes of section \ref{sec:BH} correspond to a subcase of the AdS$_2\times \Sigma$ solutions of section \ref{sec:AdS2solutions}.  Finally, in appendix \ref{app:fourcharges} we give the local form of an accelerating black hole solution in the STU model with four independent magnetic charges,  and we make a conjecture about its near horizon limit in the supersymmetric case.

\section{AdS$_2$ spindles in $D=4$ gauged supergravity}\label{sec:AdS2solutions}

In this section we present supersymmetric  AdS$_2\times \Sigma$ solutions of $D=4$,  $\mathcal{N}=4$ gauged supergravity,  where $\Sigma=\mathbb{WCP}^1_{[n_-,n_+]}$ is a spindle, parametrized by arbitrary coprime positive integers $n_{\pm}$.

\subsection{The supergravity model}\label{sec:model}

In the main part of the paper we will be interested in constructing black hole solutions, and/or their near horizon limits, to $D=4$, $\mathcal{N}=4$ gauged supergravity. This theory can also be described, in the language of $D=4$,  $\mathcal{N}=2$ supergravity \cite{deWit:1984wbb},  as a theory with no hypermultiplets and one vector multiplet,  with prepotential $F=-\ii\,X^0 X^1$ and electric Fayet-Iliopoulos gauging.  Yet 
another viewpoint is that it is a truncation  of the STU model \cite{Cvetic:1999xp},  where the four Abelian gauge fields are set pairwise equal and two of the complex scalars are identified.  Introducing the axio-dilaton 
\begin{align}
z\, = \, \frac{X^1}{X^0}\, = \, \ex^{-\xi}+\ii\,\chi\,,
\end{align}
we can write the bosonic action of the theory as
\begin{align}\label{actionU(1)^2}
\begin{split}
S\, = \, \frac{1}{16 \pi G_{(4)}} \int  &\left[(R-g^2\, \mathcal{V}) \star 1-\frac{1}{2}\mathrm{d} \xi \wedge  \star \, \mathrm{d} \xi-\frac{1}{2} \mathrm{e}^{2 \xi} \, \mathrm{d} \chi \wedge  \star\, \mathrm{d} \chi- \mathrm{e}^{-\xi} F_{2} \wedge  \star  F_{2}\right.\\
&\left.\,\,+ \,\chi\, F_{2} \wedge F_{2}-\frac{1}{1+\chi^{2} \mathrm{e}^{2 \xi}}\left(\mathrm{e}^{\xi} F_{1} \wedge  \star F_{1}+\chi\,\mathrm{e}^{2 \xi}  F_{1} \wedge F_{1}\right)\right]\,,
\end{split}
\end{align}
where $F_i=\diff A_i$, $i=1,2$, and the scalar potential $\mathcal{V}$ is given by
\begin{align}
\mathcal{V}\, = \, -\left(4+2 \cosh \xi+\mathrm{e}^{\xi} \chi^{2}\right)\,.
\end{align}
We shall henceforth set $g=1$, so that in the AdS$_4$ vacuum of the theory there is an 
effective cosmological constant $\Lambda=-3$. 
We also remark that minimal $D=4$, $\mathcal{N}=2$ gauged supergravity is obtained via the consistent truncation  
$\xi=0=\chi$, $A_1=A_2=A$. 

Although we shall not consider the fermionic completion of the action \eqref{actionU(1)^2},  it will be important to consider the supersymmetry variations of the gravitini and gaugini of this theory,  which must vanish for bosonic backgrounds that preserve some amount of supersymmetry.  While it is customary to formulate $D=4$,  $\mathcal{N}=2$ supergravity in terms of Weyl fermions,  we follow \cite{Cacciatori:2008ek} and combine them into complex Dirac fermions: a gravitino $\psi_{\mu}$,  a dilatino $\lambda$ and a supersymmetry parameter $\epsilon$.  In terms of these,  the Killing spinor equations (KSEs) can be written as\footnote{We have here corrected some typographical errors appearing in \cite{Cacciatori:2008ek}.}
\begin{align}\label{KSEs}
\begin{split}
\delta \psi_{\mu}\,  = \, &\left[\nabla_{\mu}-\frac{\ii}{2}(A_1+A_2)_\mu+\frac{\ii}{4}\ex^{\xi}\,{\partial}_{\mu}\chi\,\gamma_5+\frac{1}{4}\left(\ex^{\xi/2}+\ex^{-\xi/2}\right)\,\gamma_{\mu}+\frac{\ii}{4}\chi\,\ex^{\xi/2}\,\gamma_{\mu}\,\gamma_5\right.\\
&\left.+\frac{\ii}{8}\left(\frac{\ex^{\xi/2}}{1+\chi^2\,\ex^{2\xi}}\slashed{F}_1+\ex^{-\xi/2}\,\slashed{F}_2\right)\,\gamma_{\mu}-\frac{1}{8}\frac{\chi\,\ex^{3\xi/2}}{1+\chi^2\,\ex^{2\xi}}\,\slashed{F}_1\,\gamma_{\mu}\,\gamma_5\right]\,\epsilon \, = \, 0 \,,\\
\delta \lambda\,  = \, &\left[\ii\,\ex^{-\xi}\, \slashed{\partial}\xi-\frac{\ex^{-\xi}}{2}\left(\frac{\ex^{\xi/2}}{1+\chi^2\,\ex^{2\xi}}\,\slashed{F}_1-\ex^{-\xi/2}\,\slashed{F}_2\right)-\ii\,\ex^{-\xi}\left(\ex^{\xi/2}-\ex^{-\xi/2}\right)\right.\\
&\left. +\left(\slashed{\partial}\chi+\frac{\ii}{2}\frac{\chi\,\ex^{\xi/2}}{1+\chi^2\,\ex^{2\xi}}\,\slashed{F}_1+\chi\,\ex^{-\xi/2}\right)\,\gamma_5\right]\,\epsilon \, = \, 0\,.
\end{split}
\end{align}
Being a truncation of the maximal $D=4$,  $\mathcal{N}=8$ gauged supergravity,  all (supersymmetric) solutions of this theory can be uplifted on $S^7$ to (supersymmetric) solutions of $D=11$ supergravity.  The details of the uplift for this specific truncation can be found in \cite{Azizi:2016noi}. We shall discuss uplifting of the metric in section \ref{sec:uplift}, where global regularity of the 
$D=11$ solutions will require quantization of the magnetic charges of the $D=4$ solutions.

\subsection{Local AdS$_2$ solutions from an ansatz}

In this section we present a new class of rotating,  dyonically charged AdS$_2 \times \Sigma$ solutions of the $D=4$,  $\mathcal{N}=4$ supergravity model introduced in the previous subsection.  They are obtained from an ansatz,  as described in appendix \ref{app:ansatzsolution}, on which we imposed the equations of motion and the supersymmetry conditions.  We conjecture these solutions to arise as the near horizon limit of accelerating, rotating and dyonic black holes,  that are also extremal and supersymmetric.  While the full black hole metrics are not known in general,  we show in appendix \ref{app:ansatz} that our solutions reduce to known ones in the cases with purely magnetic charges \cite{Gauntlett:2006ns}, and with equal gauge fields \cite{Ferrero:2020twa}.  In both cases an accelerating black hole solution can also be written down,  as we discuss in section~\ref{sec:BH}.

The local form of the solutions is given by
\begin{align}\label{spinningspindle}
\begin{split}
\diff s^2_4&\, = \, \frac{1}{4}\lambda(y)\,\left(-\rho^2\,\diff\tau^2+\frac{\diff\rho^2}{\rho^2}\right)+\frac{\lambda(y)}{q(y)}\,\diff y^2+\frac{q(y)}{4\,\lambda(y)}\,(\diff z+\jp\,\rho\,\diff \tau)^2\,,\\
A_i&\, = \, \frac{h_i(y)}{\lambda(y)}\,(\diff z+\jp\,\rho\,\diff \tau)\,, \quad
\ex^{\xi}\, = \, \frac{g_1(y)}{\lambda(y)}\,, \quad
\chi\, = \, \frac{g_2(y)}{g_1(y)}\,,
\end{split}
\end{align}
where all the functions that we introduced are polynomials in $y$,  given by 
\begin{align}\label{spinningspindlefunctions}
\begin{split}
\lambda(y)\,= \, &  \, y^2+\jp^2-2c_2\,,\\
q(y)\,=\,& \, (y^2+\jp^2)^2-4(1-\jp^2+c_2)\,y^2+4c_1\,\sqrt{1-\jp^2}\,y-c_1^2+4c_2\,(c_2-\jp^2)\,,\\
h_1(y)\,=\,&\, \frac{\sqrt{1-\jp^2}}{2}\,(1-c_3)\,\lambda(y)-\frac{1}{2}\left(c_1+2\sqrt{1-\jp^2}\,\sqrt{2c_2-c_3^2\,\jp^2}\right)\,y\\
&+(2c_2-\jp^2)\,\sqrt{1-\jp^2}+\frac{1}{2}c_1\,\sqrt{2c_2-c_3^2\,\jp^2}\,,\\
h_2(y)\,=\,& \, \frac{\sqrt{1-\jp^2}}{2}\, (1+c_3)\,\lambda(y)-\frac{1}{2}\left(c_1-2\sqrt{1-\jp^2}\,\sqrt{2c_2-c_3^2\,\jp^2}\right)\,y\\
&+(2c_2-\jp^2)\,\sqrt{1-\jp^2}-\frac{1}{2}c_1\,\sqrt{2c_2-c_3^2\,\jp^2}\,,\\
g_1(y)\,=\,&\, y^2+2\,\sqrt{2c_2-c_3^2\,\jp^2}\,y+2c_2+(1-2c_3)\,\jp^2\,,\\
g_2(y)\,=\,&\, 2c_3\,\jp\,y+2\,\jp\,\sqrt{2c_2-c_3^2\,\jp^2}\,.
\end{split}
\end{align}
Note that the solution depends on the four parameters $\jp$,  $c_i$ ($i=1,2,3$),  where $\jp$ has the interpretation of a rotation parameter.  We can interpret the number of independent parameters in terms of our conjecture that this arises as the near horizon limit of a supersymmetric and extremal accelerating black hole.  One can imagine a full black hole metric with seven parameters,  representing mass,  acceleration, angular momentum and two pairs of dyonic charges.  We would then expect two constraints on the parameters to come from the supersymmetry conditions,  and one from the requirement of extremality,  resulting in a four-parameter solution, as for that described above.  One can then think of the four parameters as representing the two pairs of dyonic charges,  with mass,  acceleration and angular momentum related to them by supersymmetry and extremality.

\subsection{Killing spinors}

Let us now justify our claim that the solution \eqref{spinningspindle} is supersymmetric,  by showing  explicitly the associated Killing spinors.  To facilitate the comparison with the results in section 5.2 of \cite{Ferrero:2020twa},  we adopt the same conventions for the frame and gamma matrices,  namely we choose the orthonormal frame
\begin{align}
\begin{split}
e^0&\, =\, \frac{1}{2}\sqrt{\lambda(y)}\,\rho\,\diff\tau\,, \qquad
e^1\, = \, \frac{1}{2}\sqrt{\lambda(y)}\,\frac{\diff\rho}{\rho}\,,\\
e^2&\, = \, \sqrt{\frac{\lambda(y)}{q(y)}}\,\diff y\,, \qquad\quad \hspace{0.1cm}
e^3\, = \, \sqrt{\frac{q(y)}{4\lambda(y)}}\,\left(\diff z+\jp \,\rho\,\diff \tau\right)\,.
\end{split}
\end{align}
The four-dimensional gamma matrices are then taken to be\footnote{Explicitly,
$\gamma_0=\begin{pmatrix}
 0 &  1\\
 -1 & 0
\end{pmatrix}$, $\gamma_1=\begin{pmatrix}
 0 &  1\\
 1 & 0
\end{pmatrix}$, 
$\gamma_2=\begin{pmatrix}
 \sigma^1 & 0\\
 0 & -\sigma^1
\end{pmatrix}$, 
$\gamma_3=\begin{pmatrix}
 \sigma^2 &  0\\
 0 & -\sigma^2
\end{pmatrix}$.} 
\begin{align}\label{4dgammas}
\gamma_a&\, = \, \beta_a\otimes 1_2,  \qquad \ \  a\, = \, 0,1\,,\nn
\gamma_2&\, =\, \beta_3\otimes \sigma^1\, , \qquad 
\gamma_3\, = \, \beta_3\otimes \sigma^2\, ,
\end{align}
with the two-dimensional gamma matrices $\beta_a$ are defined by
\begin{align}\label{betaAdS}
\beta_0\, = \, \ii\,\sigma^2\, , \qquad
\beta_1\, = \, \sigma^1\, , \qquad
\beta_3\, \equiv\, \beta_0\,\beta_1\, = \, \sigma^3\,,
\end{align}
where $\sigma^i$ are the Pauli matrices.

We consider the following  Killing spinor equation (KSE) for AdS$_2$:
\begin{align}\label{AdS2KSE1}
\nabla_a \theta\, = \, \frac{\ii}{2}n\,\beta_a\, \beta_3\,\theta\, ,
\end{align}
with $n=\pm 1$. This is solved  by Majorana spinors that can be decomposed as $\theta_{1,2}=\theta^{(+)}_{1,2}+\theta^{(-)}_{1,2}$, with the
Majorana-Weyl spinors $\theta^{(\pm)}_{1,2}$ of chirality $\beta_3\, \theta^{(\pm)}_{1,2}=\pm\theta^{(\pm)}_{1,2}$,
given by \cite{Ferrero:2020twa}
\begin{align}\label{ads2spinors}
\theta^{(+)}_1& \, = \, \begin{pmatrix}
\sqrt{\rho}\\
0
\end{pmatrix}, \qquad \qquad 
\theta^{(-)}_1 \, = \,  \begin{pmatrix}
0\\
\ii\, n \sqrt{\rho}
\end{pmatrix}, \nn
\theta^{(+)}_2 & \, = \, \begin{pmatrix}
\sqrt{\rho }\,  \tau -\frac{1}{\sqrt{\rho }}\\
0
\end{pmatrix}, \quad 
\theta^{(-)}_2\, = \, \begin{pmatrix}
0\\
\ii\, n \left(\sqrt{\rho }\,  \tau +\frac{1}{\sqrt{\rho }}\right)
\end{pmatrix}\, .
\end{align}

We are finally ready to discuss the explicit Killing spinors,  which solve both equations in \eqref{KSEs},  and in analogy with \cite{Ferrero:2020twa} can be written as
\begin{align}
\label{killingspinors}
\epsilon_1&\, = \, \theta^{(+)}_1\otimes \eta_1+\theta^{(-)}_1\otimes \eta_2\, ,\nn
\epsilon_2&\, = \, \theta^{(+)}_2\otimes \eta_1+\theta^{(-)}_2\otimes \eta_2\, ,
\end{align}
where $\eta_{1,2}$ are two
two-dimensional spinors, given by
\begin{align}\label{M2spinors}
\begin{split}
\eta_1 & \, = \, 
\begin{pmatrix}
\ex^{-\tfrac{\ii}{2}\arctan\left(\frac{g_1'(y)}{2\jp\,(1-c_3)}\right)}\,\frac{q_+(y)^{1/2}}{\lambda(y)^{1/4}}\\
\ii\,\ex^{\tfrac{\ii}{2}\arctan\left(\frac{g_1'(y)}{2\jp\,(1-c_3)}\right)}\,\frac{q_-(y)^{1/2}}{\lambda(y)^{1/4}}
\end{pmatrix}\, , \\
\eta_2 & \, = \, n\,\ex^{\ii\,\arccos \jp}\,
\begin{pmatrix}
\ii\,\ex^{\tfrac{\ii}{2}\arctan\left(\frac{g_1'(y)}{2\jp\,(1-c_3)}\right)}\,\frac{q_+(y)^{1/2}}{\lambda(y)^{1/4}}\\
-\ex^{-\tfrac{\ii}{2}\arctan\left(\frac{g_1'(y)}{2\jp\,(1-c_3)}\right)}\,\frac{q_-(y)^{1/2}}{\lambda(y)^{1/4}}
\end{pmatrix}\, .
\end{split}
\end{align}
Here $g'_1(y)=\tfrac{\diff}{\diff y}g_1(y)$ and the functions $g_1(y)$ and $\lambda(y)$ were introduced in \eqref{spinningspindlefunctions}.  We have also defined
\begin{align}
q_{\pm}(y) \, \equiv \, \lambda(y)\pm \left(c_1-2\sqrt{1-\jp^2}\,y\right)\,,
\end{align}
which satisfy $q(y)=q_+(y)\,q_-(y)$.  One can check that in the limit $c_2=0=c_3$ the two-dimensional spinors \eqref{M2spinors} agree with those of \cite{Ferrero:2020twa},  after setting $c_1=\mathtt{a}$: as discussed in appendix \ref{app:minimal},  this configuration of parameters is such that \eqref{spinningspindle} is a solution of minimal gauged supergravity.

Let us conclude this subsection with some technical remarks about spinors and gamma matrices.  First, we note that the KSE \eqref{AdS2KSE1} that we have used for AdS$_2$,  which is the same used in \cite{Ferrero:2020twa},  
is a version existing only in even-dimensional AdS spaces.
However,  with a simple change of basis one can easily show that \eqref{AdS2KSE1} is equivalent to the more standard KSE\footnote{Following \cite{Fujii:1985bg}, let us consider an AdS spacetime of even dimension,  with gamma matrices $\Gamma_A$ and chirality matrix $\Gamma_*$.  If one starts with the usual KSE valid for AdS spaces of any dimension,  namely
\begin{align}
\nabla_A\,\psi\, = \, \frac{n}{2}\,\Gamma_A\,\psi\,,
\end{align}
changing basis with
\begin{align}
\Gamma_A\, = \, \ii\,\widehat{\Gamma}_A\,\Gamma_*\, = \, S\,\widehat{\Gamma}_A\,S^{-1}\,, \qquad S\, = \, \sqrt{\Gamma_*}\, ,
\end{align}
then leads to the equivalent KSE
\begin{align}
\nabla_A\,\psi\, = \, \frac{\ii}{2}\,n\,\widehat{\Gamma}_A\,\psi\,.
\end{align}
}
\begin{align}\label{AdS2KSE2}
\nabla_a\,\theta\, = \, \frac{n}{2}\,\beta_a\,\theta\,.
\end{align}
Using the two-dimensional gamma matrices $\beta_a$ given in \eqref{betaAdS},  the two independent solutions $\widetilde{\theta}_{1,2}$ to \eqref{AdS2KSE2} can be written as $\widetilde{\theta}_{1,2}=\widetilde{\theta}^{(+)}_{1,2}+\widetilde{\theta}^{(-)}_{1,2}$, with the
Majorana-Weyl spinors $\widetilde{\theta}^{(\pm)}_{1,2}$ of chirality $\beta_3\, \widetilde{\theta}^{(\pm)}_{1,2}=\pm\widetilde{\theta}^{(\pm)}_{1,2}$,
given by
\begin{align}\label{ads2spinorsnew}
\widetilde{\theta}^{(+)}_1& \, = \, \begin{pmatrix}
\sqrt{\rho}\\
0
\end{pmatrix}, \qquad \qquad 
\widetilde{\theta}^{(-)}_1 \, = \,  \begin{pmatrix}
0\\
n \sqrt{\rho}
\end{pmatrix}, \nn
\widetilde{\theta}^{(+)}_2 & \, = \, \begin{pmatrix}
\sqrt{\rho }\,  \tau -\frac{1}{\sqrt{\rho }}\\
0
\end{pmatrix}, \quad 
\widetilde{\theta}^{(-)}_2\, = \, \begin{pmatrix}
0\\
 n \left(\sqrt{\rho }\,  \tau +\frac{1}{\sqrt{\rho }}\right)
\end{pmatrix}\, .
\end{align}
Equivalently,  we can avoid giving the explicit expressions and write
\begin{align}
\widetilde{\theta}_i^{(\pm)}\, = \, A\,\theta_i^{(\pm)}\,,\qquad 
A\, = \, \left(\sqrt{\sigma^3}\right)^{-1} \, = \, 
\begin{pmatrix}
1&0\\
0&-\ii
\end{pmatrix}\,,\qquad
(i=1,2)\,.
\end{align}

The other comment is that one may expect that, at least in the non-rotating case (when the spacetime is a warped product of AdS$_2 \times \Sigma$, rather than a fibration),  it should be possible to write the four-dimensional spinors as tensor products of a single AdS$_2$ spinor and a single spinor on $\Sigma$,  rather than as a sum of two tensor products as in \eqref{killingspinors},  whose structure does not change even taking $\mathtt{j}\to 0$. With this in mind,  let us introduce a new set of four-dimensional gamma matrices $\widetilde{\gamma}_{\mu}$,  given by\footnote{Note that $\widetilde{\gamma}_{\mu}$ and ${\gamma}_{\mu}$ are of course related by a change of basis,  which reads
\begin{align}
\widetilde{\gamma}_{\mu}\, = \, M\,\gamma_{\mu}\,M^{-1}\,, \qquad M\, = \, \text{diag}(1,1,1,-1)\,.
\end{align}
}
\begin{align}\label{new4dgammas}
\widetilde{\gamma}_a&\, = \, \beta_a\otimes \beta_3,  \qquad \ \  a\, = \, 0,1\,,\nn
\widetilde{\gamma}_2&\, =\, 1_2\otimes \sigma^1\, , \qquad 
\widetilde{\gamma}_3\, = \, 1_2\otimes \sigma^2\, ,
\end{align}
where notice that with respect to \eqref{4dgammas} we have only swapped the positions of $\beta_3$ and $1_2$.  If we solve again the KSE \eqref{KSEs} using these gamma matrices,  we find two independent Killing spinors $\widetilde{\epsilon}_{1,2}$ given by
\begin{align}
\begin{split}
\widetilde{\epsilon}_{1}\, = \, \widetilde{\theta}_1^{(+)}\otimes \widetilde{\eta}_1+\widetilde{\theta}_1^{(-)}\otimes \widetilde{\eta}_2\,,\\
\widetilde{\epsilon}_{2}\, = \, \widetilde{\theta}_2^{(+)}\otimes \widetilde{\eta}_1+\widetilde{\theta}_2^{(-)}\otimes \widetilde{\eta}_2\,,
\end{split}
\end{align}
with
\begin{align}
\widetilde{\eta}_1\, = \, \eta_1\,, \quad
\widetilde{\eta}_2\, = \, -\ii\,\sigma^3\,\eta_2\,.
\end{align}
The new spinors have the property that,  in the non-rotating case $\mathtt{j}=0$,  
\begin{align}
\widetilde{\eta}\, \equiv \, \widetilde{\eta}_1\, = \, \widetilde{\eta}_2\,, \qquad (\mathtt{j}\, = \, 0)\,,
\end{align}
so that we can write
\begin{align}\label{newkillingspinors}
\widetilde{\epsilon}_{i}\, = \, \widetilde{\theta}_{i}\otimes\widetilde{\eta}\,, \qquad (i\, =\, 1 ,2)\qquad (\mathtt{j}\, = \, 0)\,,
\end{align}
which is the product structure we were looking for.  We stress that we were able to write the four-dimensional spinors as a single tensor product only in the non-rotating case,  while it seems that when rotation is present this is not possible.  It is reasonable to believe that this is due to the fact that for $\mathtt{j}\neq 0$ $\Sigma$ is fibred over AdS$_2$,  so the spacetime itself is not a product manifold.

Finally,  we conclude with some comments about the counting of supercharges.  As we have just discussed,  we have a solution to $D=4$ supergravity which admits two independent Dirac Killing spinors,  given by \eqref{killingspinors} (or equivalently \eqref{newkillingspinors}).  This is equivalent to four Majorana,  or four Weyl spinors,  hence the solution can be described as $\tfrac{1}{2}-$BPS from the point of view of $D=4$, $\mathcal{N}=2$ supergravity,  or $\tfrac{1}{4}-$BPS from the point of view of $D=4$, $\mathcal{N}=4$ supergravity.  In the dual $d=1$ superconformal quantum mechanics (SCQM),  the complex spinor $\epsilon_1$ gives two real Poincar\'e supercharges,  while $\epsilon_2$ gives two real conformal supercharges.  Thus,  the SCQM has $\mathcal{N}=2$ supersymmetry in one dimension,  since in the field theory counting one usually includes only Poincar\'e supercharges,  with superalgebra $\mathfrak{su}(1,1|1)$.

\subsection{Global analysis}

We would now like to determine conditions on the parameters $\jp$ and $c_i$ ($i=1,2,3$) such that the two-dimensional metric 
\begin{align}\label{spindlemetric}
\diff s^2_{\Sigma}\, = \, \frac{\lambda(y)}{q(y)}\,\diff y^2+\frac{q(y)}{4\,\lambda(y)}\,\diff z^2\,,
\end{align}
obtained from \eqref{spinningspindle} on slices of constant $\tau$ and $\rho$,  is a smooth orbifold metric on a spindle $\Sigma=\mathbb{WCP}^1_{[n_-,n_+]}$.  Clearly,  we want $\lambda(y)>0$ and $q(y)\geq 0$,  which is also enough to guarantee the correct signature of the metric \eqref{spinningspindle}.  For \eqref{spindlemetric} to be a metric on a compact space,  we also want to take $y\in [y_a,y_b]$,  with $y_a<y_b$ two roots of $q(y)=0$,  such that $q(y)>0$ for $y\in (y_a,y_b)$.  Since the coefficient of $y^4$ in $q(y)$ is positive,  this is only possible if there are four single\footnote{If there is a double root,  that is necessarily either $y_a$ or $y_b$,  but then \eqref{spindlemetric} would not yield  a complete metric on a compact space.} real roots,  and $y_{a,b}$ are taken to be the middle two roots.

A sufficient condition for $\lambda(y)$ to be positive is that it has no real roots,  which is the case for $c_2<\frac{\jp^2}{2}$.  As for the roots of $q(y)$,  they admit a simple expression as
\begin{align}\label{rootsq}
\begin{split}
y_1&\, = \, -\sqrt{1-\jp^2}-\sqrt{1+c_1+2c_2-2\jp^2}\,, \\
y_2& \, = \, -\sqrt{1-\jp^2}+\sqrt{1+c_1+2c_2-2\jp^2}\,, \\
y_3&\, = \, +\sqrt{1-\jp^2}-\sqrt{1-c_1+2c_2-2\jp^2}\,, \\
y_4& \, = \, +\sqrt{1-\jp^2}+\sqrt{1-c_1+2c_2-2\jp^2}\,,
\end{split}
\end{align}
and note that for at least two of the roots to be real we need $\jp\in [-1,1]$.  Since the sign of $\jp$ can be reabsorbed with a change of the sign of $\tau$,  we are actually free to set $\jp \in [0,1]$.  We further note that $y_{1,2}$ are real and distinct for $c_1> -f(\jp,c_2)$,  while $y_{3,4}$ are real and distinct for $c_1< f(\jp,c_2)$,  with $f(\jp,c_2)=1+2c_2-2\jp^2$.  Thus,  a necessary condition to have four distinct real roots is that $f(\jp,c_2)> 0$,  which leads to the constraint $c_2> \jp^2-\frac{1}{2}$.  Note that in this case we also have $y_1<y_2<y_3<y_4$,  so we must set $a=2$, $b=3$ and take $y\in [y_2,y_3]$.  Finally,  we also note that the dilaton $\ex^{\xi}$ should be positive for $\xi$ to be real.  Its denominator $\lambda(y)$ is positive in the ranges discussed above,  while the numerator $g_1(y)$ is a polynomial in $y$ of degree two which is always positive since it has a negative discriminant,  given by $-4\,(1-c_3)^2\,\mathtt{j}^2$.  Finally,  we should also take $c_2$ such that the square root $\sqrt{2\,c_2-c_3^2\,\mathtt{j}^2}$ appearing in \eqref{spinningspindlefunctions} is real,  which requires $ 2 c_2\ge c_3^2\,\mathtt{j}^2$.  This gives a non-empty intersection with the other conditions (in particular $c_2<\tfrac{1}{2}\jp^2$) only if $|c_3|<1$. 

To summarize,  we have shown that when\footnote{Note that from \eqref{existenceconditions} it seems that taking $\jp=0$ also forces $c_2=0$.  This is,  however,  not the case,  since the correct way to turn off the rotation parameter is that of taking a limit $c_3\to \infty$ and $\jp \to 0$,  with constant product $c_3\,\jp$.  We discuss this in appendix \ref{app:dyonicnonrot}.}
\begin{align}\label{existenceconditions}
0\, \le \,  \jp \, \le \, 1\,, \quad \ 
\text{max}\left(\jp^2-\tfrac{1}{2},\tfrac{1}{2} c_3^2\,\mathtt{j}^2\right)\, < \,  c_2 \, < \,  \tfrac{1}{2}\jp^2\,,\quad \ 
|c_1|\, <\, 1+2c_2-2\jp^2\, , \quad |c_3|<1\,,
\end{align}
we can take $y\in [y_2,y_3]$,  with $q(y)\geq 0$ and $\lambda(y)>0$ in that interval.  We shall from now on assume that these conditions hold,  and study the global regularity of \eqref{spindlemetric} under this assumption.  

Let us then consider the behaviour of the metric \eqref{spindlemetric} near the poles $y_{a,b}$.  For any $y_i$ such that $q(y_i)=0$,  setting $y=\frac{r^2}{4}+y_i$ we find
\begin{align}
\diff s^2_{\Sigma} \, \simeq \, \frac{\lambda(y_i)}{q'(y_i)}\,\left(\diff r^2+r^2\frac{q'(y_i)^2}{16\lambda(y_i)^2}\,\diff z^2\right)\,.
\end{align}
Then,  \eqref{spindlemetric} is a smooth metric\footnote{In the orbifold sense: the metric is regular everywhere except for the poles $y=y_{2,3}$,  where there are conical deficit angles $2\pi\left(1-\frac{1}{n_{\pm}}\right)$.} on $\mathbb{WCP}^1_{[n_-,n_+]}$ if 
\begin{align}\label{quantisespindle}
\frac{q'(y_2)}{4\,\lambda(y_2)}\,\Delta z = \, \frac{2\pi}{n_+}\,,  \qquad
-\frac{q'(y_3)}{4\,\lambda(y_3)}\, \Delta z= \, \frac{2\pi}{n_-}\,,
\end{align}
with $n_{\pm}$ coprime positive integers. Notice here that $\lambda>0$, while $q'(y_2)>0$ and $q'(y_3)<0$, which determines the signs in 
\eqref{quantisespindle}. 
These equations are solved by
\begin{align}\label{quantiseparameters}
c_1\, = \, \frac{(n_-^2-n_+^2)\,(1+2c_2-2\jp^2)}{n_-^2+n_+^2}\,, \qquad
\Delta z\, = \, \frac{\sqrt{2}\,\sqrt{n_-^2+n_+^2}}{n_-\,n_+\,\sqrt{1+2c_2-2\jp^2}}\,\pi\,.
\end{align}
Using these conditions, and using the expression
\begin{align}
\sqrt{g_{\Sigma}}\,R_{\Sigma}\, = \, \frac{\diff}{\diff y}\frac{q(y)\,\lambda'(y)-q'(y)\,\lambda(y)}{2\lambda(y)^2}\,,
\end{align}
for the Ricci scalar of the metric \eqref{spindlemetric},  we can also check that the Euler number
\begin{align}\label{Euler}
\chi(\Sigma)\, = \, \frac{1}{4\pi}\int_{\Sigma}\,R_{\Sigma}\,\text{vol}_{\Sigma}\, = \, \frac{n_-+n_+}{n_-\,n_+}\,,
\end{align}
takes the correct value for the spindle.  Note that the last condition in \eqref{existenceconditions} is trivially satisfied for all values of $n_{\pm}$ due to the constraint \eqref{quantiseparameters}.

\subsection{Conserved charges and entropy of AdS$_2$ spinning spindles}\label{sec:charges}

Having established the conditions for which \eqref{quantisespindle} is a metric on a spindle, in this section we shall compute the conserved charges associated with the (conjectural) black hole of which \eqref{spinningspindle} represents the near horizon limit,  as well as its entropy.  
In the full black hole solution, these conserved charges would usually be defined as integrals 
over a constant time surface $\Sigma_\infty$ at infinity, the integrand being constructed from an 
appropriately conserved current.
 However, at least for the electric and magnetic charges and 
angular momentum, using Stokes' Theorem we may equivalently evaluate these quantities 
as integrals over the horizon $\Sigma$, which may then be computed in the near horizon solution, following \cite{Ferrero:2020twa}.

First,  we define the magnetic charges to be
\begin{align}\label{magflux}
P_i\, \equiv \, \frac{1}{2\pi}\int_{\Sigma}F_i\, .
\end{align}
Since $\diff F_i=0$, these charges will be equal to $\frac{1}{2\pi}\int_{\Sigma_\infty}F_i$ for any solution 
in which the horizon $\Sigma$ is homologous to a spacelike surface $\Sigma_\infty$ at infinity. 
After a computation we find
\begin{align}\label{Ps}
\begin{split}
P_1+P_2\,&  = \, \frac{n_--n_+}{n_-\,n_+}\,  \, \equiv\, 4G_{(4)} Q_m\, , \\
P_1-P_2\, & =\, -2\sqrt{2c_2-c_3^2\,\jp^2}\,\frac{\Delta z}{2\pi}\,,
\end{split}
\end{align}
where $Q_m$ was first introduced in \cite{Ferrero:2020twa}, and we have 
included a factor of $G_{(4)}$ in its definition, as in \cite{Cassani:2021dwa}. 
The first equation in \eqref{Ps} gives the same ``anti-topological twist'' 
encountered in \cite{Ferrero:2020laf, Ferrero:2020twa}, where 
the total magnetic flux is $P_1+P_2=(n_--n_+)/n_-n_+$. 
The nomenclature anti-topological twist was introduced in \cite{Ferrero:2021wvk}, 
due to the relative minus sign $(n_--n_+)/n_-n_+$ in this expression for the total flux. 
This may be contrasted with the  
Euler number of the spindle $\chi(\Sigma)$ given  by \eqref{Euler}. 
The latter would be the total magnetic flux $P_1+P_2$ if 
supersymmetry was realized by a topological twist, appropriately identifying 
the spin connection on $\Sigma$ with the R-symmetry gauge fields, so that 
the Killing spinor is constant. However,
for the anti-topological twist here the spinors are sections 
of non-trivial bundles over $\Sigma$, as in \cite{Ferrero:2020twa}, and so certainly not constant.

To define the electric charges,  we notice that while in general $\diff \star F_i \neq 0$,  the two-forms
\begin{align}
\mathcal{F}_1\, \equiv \, \frac{\ex^{\xi}}{1+\chi^2\,\ex^{2\xi}}\left(\star F_1+\chi\,\ex^{\xi}\,F_1\right)\,,  \qquad
\mathcal{F}_2\, \equiv  \, \ex^{-\xi}\,\star F_2-\chi\,F_2\,,
\end{align}
are closed by virtue of the equations of motion.  We thus define
\begin{align}
Q_i\, \equiv \, -\frac{1}{2\pi}\int_{\Sigma}\mathcal{F}_i\,,
\end{align}
which by a similar comment to that above will be equal to the corresponding integrals 
evaluated on $\Sigma_\infty$. We find
\begin{align}\label{Qs}
\begin{split}
Q_1+Q_2\, & = \, 2\jp\,\frac{\Delta z}{2\pi}\, \, \equiv \, 4 G_{(4)}Q_e\, , \\
Q_1-Q_2\, & = \, -c_3\,(Q_1+Q_2)\,.
\end{split}
\end{align}
Again, in the first equation we have defined the total electric charge $Q_e$, which 
coincides with the quantity defined in \cite{Ferrero:2020twa,Cassani:2021dwa}.

Even without knowing the full black hole metric of which \eqref{spinningspindle} is the near horizon limit,  we can still compute its entropy using the Bekenstein-Hawking formula
\begin{align}
\begin{split}
S_{BH}& \, = \, \frac{\text{Area}}{4G_{(4)}}\, = \, \frac{1}{4G_{(4)}}\frac{y_3-y_2}{2}\,\Delta z \\
& \, = \, \frac{\pi}{4G_{(4)}}\,\left(\frac{\sqrt{2(n_-^2+n_+^2)\,(1-\jp^2)}}{n_-\,n_+\,\sqrt{1+2c_2-2\jp^2}}-\frac{n_-+n_+}{n_-\,n_+}\right)\,.
\end{split}
\end{align}
In terms of the two pairs of dyonic charges of this solution,  the entropy can be also expressed as
\begin{align}\label{SBH}
S_{BH}\, = \, \frac{\pi}{4G_{(4)}}\left[-\chi(\Sigma)+\sqrt{\chi(\Sigma)^2+4(P_1 P_2+Q_1 Q_2)}\right]\, .
\end{align}
We shall derive this formula in a very different way in section \ref{sec:conjectures}, where we will 
also discuss various special cases which have previously appeared in the literature.

Another physical quantity that can be computed for a rotating black hole is its angular momentum.  Since the metric of the full black hole is not known,  we shall adopt (a suitably modified version of) the prescription of \cite{Ferrero:2020twa},  where the angular momentum is defined as a sort of Page charge. To define this, we first introduce an angle $\varphi=\frac{2\pi}{\Delta z}\,z$ and a Killing vector $k=\partial_{\varphi}$,  in terms of which the angular momentum can then be expressed as
\begin{align}\label{Jdef}
J(A_1,\,A_2)\, = \, \frac{1}{16\pi}\left[\int_{\Sigma}\star\, \diff k+2\,(k\cdot A_1)\,\mathcal{F}_1+2\,(k\cdot A_2)\,\mathcal{F}_2\right]\,.
\end{align}
Although the integrand here is not a closed form, so that this doesn't immediately lead to a conserved quantity, 
one can verify that $k\lrcorner\, \diff$ applied to the integrand is zero. Assuming that the horizon $\Sigma$ of the near horizon black hole solution
and the corresponding copy of this surface $\Sigma_\infty$ on the conformal boundary are the two boundary components of a $k$-invariant three-manifold, 
as one would expect for the black hole solution, 
it follows from Stokes' Theorem that \eqref{Jdef} takes the same value integrated over either $\Sigma$ or $\Sigma_\infty$. 
However, being a type of Page charge, this angular momentum is not gauge invariant. 
We will evaluate it in the gauge given in \eqref{spinningspindle},  which is natural from the point of view of a near horizon solution as it is invariant under the (twisted) isometries of AdS$_2$ (see \cite{Ferrero:2020twa} for more details).  We refer to the value of the angular momentum computed in this gauge as $J_{AdS_2}$,  and we find
\begin{align}
J_{AdS_2}\, = \, \frac{1-c_3^2}{4}\,\jp\,\sqrt{1-\jp^2}\,\left(\frac{\Delta z}{2\pi}\right)^2\,,
\end{align}
which agrees with the result found in \cite{Ferrero:2020twa} when $c_3=0$.  Note that we can also write 
\begin{align}\label{JAdS2}
J_{AdS_2}\, = \, \frac{Q_1Q_2}{4(Q_1+Q_2)}\sqrt{\chi(\Sigma)^2+4(P_1P_2+Q_1Q_2)}\,,
\end{align}
and thus the entropy can be rewritten as
\begin{align}\label{SBHJ}
S_{BH}\, = \, \frac{\pi}{4G_{(4)}}\left[\frac{4 (Q_1+Q_2)}{Q_1 Q_2}\,J_{AdS_2}-\chi(\Sigma)\right]\,.
\end{align}

\subsection{Uplift to $D=11$}\label{sec:uplift}

As already commented at the end of section \ref{sec:model}, the solutions we have constructed can automatically be uplifted \emph{locally} on $S^7$ to supersymmetric solutions of $D=11$ supergravity.  The relevant uplifting formulas are given in \cite{Azizi:2016noi}. In this section 
we briefly comment on the conditions required for global regularity of these $D=11$ solutions.

The uplifted $D=11$ metric is \cite{Azizi:2016noi} given by
\begin{align}\label{uplift}
\begin{split}
L^{-2}\, \diff s^2_{11} & \, = \,  (UV)^{1/3}\diff s^2_4 + 4(UV)^{1/3}\Big\{ \diff \eta^2 + 
\frac{\cos^2\eta}{4V}\big[\diff\theta_1^2 + \sin^2\theta_1 \diff\phi_1^2  \\ 
& \quad \ + (\diff \psi_1 + \cos\theta_1 \diff \phi_1  - A_1)^2\big] + 
\frac{\sin^2\eta}{4U}\big[\diff\theta_2^2 + \sin^2\theta_2 \diff\phi_2^2 \\ 
& \quad \ + (\diff \psi_2 + \cos\theta_2 \diff\phi_2 - A_2)^2\big] \Big\}\, .
\end{split}
\end{align}
Here we have introduced an overall constant length scale $L>0$, and have defined  the functions
\begin{align}
U \, \equiv \, (\ex^{-\xi} +\chi^2\, \ex^{\xi})\sin^2\eta + \cos^2\eta\, , \qquad V \, \equiv \, \ex^\xi \cos^2\eta + \sin^2\eta\, .
\end{align}
Here the metric in curly brackets is a metric on $S^7$, where one views $S^7\subset \C^2\oplus\C^2$ as unit sphere, 
with the metrics in square brackets being metrics on the two copies of $S^3\subset \C^2$. 
It follows that $\theta_i\in [0,\pi]$, $\eta\in[0,\frac{\pi}{2}]$, while $\phi_i$ have period $2\pi$ 
and $\psi_i$ have period $4\pi$, $i=1,2$. The gauge fields $A_i$ then fibre 
the two three-spheres over the $D=4$ spacetime, effectively gauging 
the Hopf $U(1)$ isometry of each $S^3$. The formula for the $D=11$ four-form flux $G$
is rather more involved, and can be found in \cite{Azizi:2016noi}.

For the spinning spindle solutions \eqref{spinningspindle} we have constructed the gauge 
fields are not in general globally defined one-forms on AdS$_2\times \Sigma$, as must
be the case since the magnetic fluxes in \eqref{magflux} are generically non-zero. 
On the other hand, these gauge fields fibre the internal $S^7$ over this spacetime 
via \eqref{uplift}, and this will lead to a globally well-defined $D=11$ spacetime 
only if the $P_i$ satisfy certain quantization conditions. Specifically, as in \cite{Ferrero:2020twa}  
this requires
\begin{align}\label{Pquantize}
P_i \, = \, \frac{2\,\mathtt{p}_i}{n_-n_+}\, , 
\end{align}
where $\mathtt{p}_i\in\Z$ are integers coprime to $n_\pm$.\footnote{The factor of $2$ in \eqref{Pquantize} arises because 
$\psi_i$ has period $4\pi$, rather than the canonical $2\pi$.} Imposing \eqref{Pquantize} the $D=11$ spacetime 
is then the total space of an $S^7$ fibration over AdS$_2\times \Sigma$, with this total space being 
free from orbifold singularities.

The $D=11$ solution can be understood as the near horizon limit of $N$ M2-branes wrapped on
the spindle $\Sigma$, where the flux number $N$ is  defined by
\begin{align}
N\, =\, \frac{1}{(2\pi\ell_p)^6}\int_{S^7} \star_{11} G +  \frac{1}{2}C\wedge G\, .
\end{align}
Here  the $S^7$ is a copy of the fibre, at any point in the $D=4$ spacetime.
We find that in turn this fixes the constant $L$ via
\begin{align}
\frac{L^6}{(2\pi \ell_p)^6} \, = \, \frac{N}{128\pi^4}\, ,
\end{align}
while the $D=4$ Newton constant is
\begin{align}
\frac{1}{G_{(4)}} \, = \, \frac{2\sqrt{2}}{3} N^{3/2}\, .
\end{align}

\section{Magnetic spindles from accelerating black holes}\label{sec:BH}

In this section we consider a new family of accelerating black holes with two magnetic charges in AdS$_4$,  constructed via electromagnetic duality from the analogous electrically charged solutions of \cite{Lu:2014sza}.  We consider the case when such black holes are supersymmetric and extremal,  and we show that their near horizon limit corresponds to the limit of vanishing rotation and electric charges of the solutions presented in section \ref{sec:AdS2solutions}.  Finally we show that,  in analogy with the results of \cite{Ferrero:2020twa},  the conformal boundary of these black holes has a singularity where it intersects an acceleration horizon.

\subsection{Multi-charge accelerating black holes}

Our starting point is the class of electrically charged,  accelerating black holes presented in \cite{Lu:2014sza}.  Using electromagnetic duality of the theory with no axions,  one can immediately obtain an analogous solution where only magnetic charges are present.  Choosing a convenient parametrization,  we can express the corresponding solution as 
\begin{align} \label{magneticBH}
\begin{split}
\diff s^2&\, = \, \frac{1}{H^2}\left[ -Q\frac{\Gamma}{\Sigma}\,\diff t^2+\frac{\Sigma\Gamma}{Q}\diff r^2+\frac{\Sigma\Gamma}{P}\,\diff \theta^2+\frac{\Sigma}{\Gamma}P\sin^2\theta\,\diff\phi^2\right]\,\,,\\
A_1&\, = \, \frac{\sqrt{1-\alpha^2\deltap^2}\,\sqrt{\sigmap+2m\deltap}}{\alpha\deltap(1+\alpha\deltap\cos\theta)}\,\diff \phi\,,\qquad
A_2\, = \, -\frac{\sqrt{1-\alpha^2\deltap^2}\,\sqrt{\sigmap-2m\deltap}}{\alpha\deltap(1-\alpha\deltap\cos\theta)}\,\diff \phi\,,\\
\ex^{\xi}&\, = \, \frac{(r-\deltap)(1+\alpha\deltap\cos\theta)}{(r+\deltap)(1-\alpha\deltap\cos\theta)}\,,\qquad
\chi\, = \, 0\,.
\end{split}
\end{align}
In the above,  we have introduced the functions 
\begin{align}\label{functionsmagneticBH}
\begin{split}
Q\, = \, &\, \Big (r^2-2m r +\sigmap-\deltap^2\Big )(1-\alpha^2 r^2) +(r^2-\deltap^2)^2\,,\\
P\, = \, &\, 1-2m \alpha \cos\theta+\alpha^2(\sigmap-\deltap^2)\cos^2\theta\,,\\
\Sigma\, = \, &\, r^2-\deltap^2\,,  \qquad
\Gamma\, = \, 1-\alpha^2 \deltap^2\cos^2\theta\,,\\
H \, = \, &\, 1-\alpha r \cos\theta\, .
\end{split}
\end{align}
The solution is fully characterized by the four parameters $m$,  $\alpha$,  $g$ and $\deltap$: the former two are related to the  mass and the acceleration,  respectively,  while the latter are related to the two magnetic charges. Note in particular that $\deltap=0$ sets the two magnetic charges to be equal,  and \eqref{magneticBH} then reduces to a solution of minimal gauged supergravity\footnote{In the potentials $A_i$ one should expand around $\delta\to 0$ and remove the singular terms with suitable gauge transformations.}.

In the above,  we take the range of the coordinate $\theta$ to be $0\le \theta \le \pi$,  and we take the parameters to be such that $P>0$ and $\Gamma>0$ in that range.  The precise range of the parameters will be discussed  in more detail when focusing on the supersymmetric black hole. To determine the range of $\phi$,  we expand the metric near the poles $\theta_-=0$ and $\theta_+=\pi$,  to find
\begin{align}
\diff s^2_{\theta,\phi}\, \approx \,  \left[\frac{\Gamma\Sigma}{H^2P^2}\right]_{\theta=\theta_{\pm}}\,\left[\diff\theta^2+(\theta-\theta_{\pm})^2\,\frac{P_{\pm}^2}{\Gamma_{\pm}^2}\,\diff\phi^2\right]\,,
\end{align}
where 
\begin{align}
P_{\pm}\, = \, P(\theta_{\pm})\, = \, 1\pm 2m\alpha+\alpha^2(\sigmap-\deltap^2)\,,\qquad
\Gamma_{\pm}\, = \, \Gamma(\theta_{\pm})\, = \, 1-\alpha^2 \deltap^2\,.
\end{align} 
Note that while $\Gamma_+=\Gamma_-$,  since $P_+\neq P_-$ it is impossible to choose a period for $\phi$ such that we obtain a smooth metric on a two-sphere, and there will always be conical deficits at the poles.  Hence,  following \cite{Ferrero:2020twa},  we quantize the deficits so that the space parametrized by the coordinates $\theta$ and $\phi$ is a spindle.  This is achieved by requiring that
\begin{align}\label{quantpolesfullBH}
\frac{P_{\pm}}{\Gamma_{\pm}}\Delta\phi\, = \, \frac{2\pi}{n_{\pm}}\,.
\end{align}
We shall not solve this equation now, but  rather in the next section when we consider the case in which the black hole is supersymmetric. 
There we  also explicitly compute the period of $\phi$,  as well as the entropy and the magnetic charges.

\subsection{Supersymmetric and extremal limit}

We now focus on the supersymmetric case of the accelerating and magnetic black holes introduced in the previous subsection.  In order for supersymmetry to be preserved there must be a non-trivial solution to the gaugino equation in \eqref{KSEs}.  The latter can be written as
\begin{align}
\delta \lambda \, = \, 0 \, = \, \mathcal{M}\,  \epsilon\,,
\end{align}
for some matrix $\mathcal{M}$,  and a necessary condition for this to admit non-trivial solutions is that $\det \mathcal{M}=0$.  The solution to the latter equation is most easily expressed in terms of $\alpha$ and a new parameter $\xp$,  using
\begin{align}\label{magneticBHsusy}
\begin{split}
m\, & = \, \frac{1-\alpha^2-\xp^2}{\alpha^3\sqrt{1-\alpha^2}}\,,  \qquad
\deltap\, = \, \frac{\xp}{\alpha \sqrt{1-\alpha^2}}\,, \\
\sigmap\, & = \, \frac{(1-\alpha^2-\xp^2)(1+\xp^2-2\alpha^2+\alpha^4)}{\alpha^4(1-\alpha^2)^2}\,.
\end{split}
\end{align}
It turns out that not only are these conditions necessary and in fact sufficient for supersymmetry to be preserved,  but the resulting solution is also extremal.  Indeed,  when the parameters take the values in \eqref{magneticBHsusy},  we can write the function $Q$ in \eqref{functionsmagneticBH} as
\begin{align}
Q\, = \, (1-\alpha^2)^2(r-r_-)^2(r-r_+)^2\,,
\end{align}
with roots
\begin{align}\label{horizonradii}
r_{\pm}\, = \, \frac{-(1-\alpha^2-\xp^2)\pm \sqrt{\xp^4-2\xp^2(1-\alpha^2)+(1-\alpha^2)^2(5-4\alpha^2)}}{2\alpha(1-\alpha^2)^{3/2}}\,,
\end{align}
corresponding to an ordinary horizon ($r_+$) and an acceleration horizon ($r_-$), respectively.

Let us from now on focus on this supersymmetric case,  and analyse the conditions for global regularity of the black hole in detail.  First,  we note that the metric functions $P$ and $\Gamma$ can be written as 
\begin{align}
\begin{split}
P&\, = \, 1-\frac{2}{\alpha^2\sqrt{1-\alpha^2}}(1-\alpha^2-\xp^2)\cos\theta+\frac{(1-\alpha^2)^3-\xp^4}{\alpha^2(1-\alpha^2)^2}\cos^2\theta\, ,\\
\Gamma&\, =\, 1-\frac{\xp^2}{1-\alpha^2}\cos^2\theta\,,
\end{split}
\end{align}
and we remind the reader that since we are taking $0\le \theta \le \pi$,  we want $P>0$ and $\Gamma>0$ in that range.  Using the symmetry that exchanges $\theta \leftrightarrow \pi-\theta$ we can always choose $\alpha>0$: reality of the metric then requires $0<\alpha<1$.  Likewise,  we can focus on $\xp>0$,  as sending $\xp \to -\xp$ simply amounts to swapping the two gauge fields.  Then,  we have that $\Gamma>0$ for $x<\sqrt{1-\alpha^2}$,  while we find that to have $P>0$ one must take $\frac{\sqrt{3}}{2}<\alpha<1$.  Note that in this range we always have $r_+>\alpha^{-1}>0$ and $r_-<0$.

A careful analysis is also required to determine the range of $r$.  We note that the conformal boundary is reached when  $H=0$,  namely for 
\begin{align}
r\, = \, \frac{1}{\alpha \cos\theta}\,,
\end{align}
which is negative for $\frac{\pi}{2}<\theta<\pi$.  Hence,  for $\theta>\pi/2$ the coordinate $r$ is allowed to take negative values,  and the allowed values of $r$ are given by 
\begin{align}
\begin{split}
&0<\theta<\pi/2\,: \quad r_+<r<(\alpha\cos\theta)^{-1}\,,\\
&\pi/2<\theta<\pi\,: \quad r>r_+ \quad \text{or} \quad r<\text{min}\left\{r_-,\,(\alpha \cos\theta)^{-1}\right\}\,.
\end{split}
\end{align}
While $Q$ is guaranteed to be positive in this range,  we also need to require that $\Sigma>0$,  which is true for $r_+>\frac{x}{\alpha\,\sqrt{1-\alpha^2}}$ and $r_-<-\frac{x}{\alpha\,\sqrt{1-\alpha^2}}$.  The latter conditions can be satisfied by further restricting the range of $\xp$ to be $0<\xp<1-\alpha^2$.  A detailed analysis of the resulting conformal boundary will be given in the next subsection.  

To summarize,  taking the parameters $\alpha$ and $\xp$ to satisfy
\begin{align}\label{parametersBH}
\frac{\sqrt{3}}{2}\, <\, \alpha \, < \, 1\,,  \qquad
0\, < \, \xp \, <\, 1-\alpha^2\,,
\end{align}
we can obtain a globally regular metric,  except for the conical deficits located at $\theta=\theta_{\pm}$.  As already explained in the previous subsection, we choose to quantize these conical deficits so that the space in the $\theta$ and $\phi$ directions is a spindle $\mathbb{WCP}^1_{[n_-,n_+]}$.  Solving explicitly the condition \eqref{quantpolesfullBH},  we find that this implies
\begin{align}
\xp\, = \, \sqrt{1-\alpha^2}\sqrt{2\, \frac{n_-+n_+}{n_--n_+}\sqrt{1-\alpha^2}-1}\,, \quad
\Delta \phi\, = \, \frac{\alpha^2\pi}{2\sqrt{1-\alpha^2}}\frac{n_--n_+}{n_-\,n_+}\,.
\end{align}

From this result we can easily compute the charges and the entropy associated with the supersymmetric black hole.   For the magnetic charges,  we find
\begin{align}\label{magneticchargesBH}
\begin{split}
P_1&\, = \, \frac{1}{2\pi}\int F_1\, = \, \frac{n_--n_+}{2n_-n_+}\left(1+\frac{1}{\sqrt{1-\alpha^2}}\sqrt{2\, \frac{n_-+n_+}{n_--n_+}\sqrt{1-\alpha^2}-1}\right)\,,\\
P_2&\, = \, \frac{1}{2\pi}\int F_2\,  = \, \frac{n_--n_+}{2n_-n_+}\left(1-\frac{1}{\sqrt{1-\alpha^2}}\sqrt{2\, \frac{n_-+n_+}{n_--n_+}\sqrt{1-\alpha^2}-1}\right)\,,
\end{split}
\end{align}
from which it immediately follows that 
\begin{align}
P_1+P_2\, = \, \frac{n_--n_+}{n_-n_+}\, .
\end{align}
Of course, this coincides with the first equation in \eqref{Ps} in the near horizon solution, and 
describes an anti-topological twist.

We can also compute the entropy of the supersymmetric black hole from the Bekenstein-Hawking formula,  which gives
 \begin{align}
 \begin{split}
S_{BH}&\, = \, \frac{\text{Area}}{4G_{(4)}} \, = \, \frac{r^2-\frac{x^2}{\alpha^2(1-\alpha^2)}}{1-\alpha^2r_+^2}\,\frac{\Delta\phi}{2G_{(4)}}\\
&\, =\, \frac{\pi}{4G_{(4)}}\Bigg[\frac{\sqrt{-2n_-n_++(3-2\alpha^2)(n_-^2+n_+^2)-2\sqrt{1-\alpha^2}(n_-^2-n_+^2)}}{\sqrt{1-\alpha^2}\, n_-n_+ } \\ 
& \qquad \qquad \qquad  -\frac{n_-+n_+}{n_-n_+}\Bigg]\,.
\end{split}
\end{align}
We can also express the entropy in terms of the physical magnetic charges \eqref{magneticchargesBH} and the Euler number of the spindle
\begin{align}
\chi(\Sigma)\,= \, \frac{n_-+n_+}{n_-\,n_+}\,,
\end{align}  
which gives
\begin{align}
S_{BH}\, = \, \frac{\pi}{4G_{(4)}}\left[-\chi(\Sigma)+\sqrt{\chi(\Sigma)^2+4P_1P_2}\right]\,.
\end{align}
This agrees with the entropy computed in \eqref{SBH} for the AdS$_2\times \Sigma$ solutions of section~\ref{sec:AdS2solutions},  in the case with no electric charges ($Q_1=0=Q_2$), and therefore vanishing angular momentum as well.  
This is not a coincidence,  but rather a consequence of the fact that the near horizon limit of the supersymmetric and extremal black hole described here gives precisely the solution \eqref{spinningspindle},  when rotation and electric charges are turned off (namely, $c_3=\jp=0$).  The details of this computation are given in appendix \ref{app:NHlimit},  while here we just mention that the AdS$_2$ solution in this limit can also be obtained from the $D=11$ solution with four magnetic charges found in \cite{Gauntlett:2006ns},  using the appropriate reduction formulas and setting the charges to be pairwise equal.  The connection with that solution is discussed in appendix \ref{app:purelymag}.

\subsection{The conformal boundary}

Let us conclude the analysis of the black hole solution \eqref{magneticBH} by studying its conformal boundary,  focusing again on the supersymmetric case since it is the relevant one for this paper.  As anticipated,  the conformal boundary corresponds to the locus  $H=0$,  which is given by 
\begin{align}
r \, = \, \frac{1}{\alpha\cos\theta}\,,
\end{align}
so that when $\theta>\pi/2$ the coordinate $r$ is allowed to take negative values.  We then follow \cite{Ferrero:2020twa} and introduce $y=1/r$,  in terms of which the conformal boundary is located at $y=\alpha\cos\theta$.  The coordinate $r$ is in principle allowed to range from the horizon $r_+$ to the conformal boundary,  so that setting $y_H=1/r_+>0$ one should have
\begin{align}
\alpha \cos\theta \, < \, y \, < \, y_H\,,
\end{align}
and since $y_H>\alpha$ in the range of parameters \eqref{parametersBH},  the ordering above always makes sense.  However,  since $r$ (and so $y$) is allowed to take negative values,  one can in principle also reach the other double root of $Q$,  namely the acceleration horizon $r_-$,  or $y_A=1/r_-<0$.  This intersects the conformal boundary when 
\begin{align}
y_A\, = \, \alpha \cos\theta,
\end{align}
which can be solved for
\begin{align}\label{boundary_vs_yA}
\begin{split}
\theta\, & = \, \theta_0 \\
& \equiv \, \arccos\left(-\frac{2(1-\alpha^2)^{3/2}}{(1-\alpha^2-\xp^2)+ \sqrt{\xp^4-2\xp^2(1-\alpha^2)+(1-\alpha^2)^2(5-4\alpha^2)}}\right)\,.
\end{split}
\end{align}
Note that for all allowed values of the parameters in \eqref{parametersBH} we have $\pi/2<\theta_0<\pi$.  This means that for $\theta>\theta_0$ one has $y_A>\alpha\cos\theta$,  and the acceleration horizon is reached {\it before} the conformal boundary,  which is then partially hidden behind this horizon.  This is exactly the same behaviour that was observed in \cite{Ferrero:2020twa}.

The explicit boundary metric can be computed directly from \eqref{magneticBH},  where we choose a particular representative of the conformal class such that the Killing vector $\partial_t$ is unit normalized.  Setting $r=(\alpha \cos\theta)^{-1}$ and using the identities
\begin{align}\label{BHidentities}
\begin{split}
0&\, = \, \alpha^4\cos^4\theta\, Q\left(\frac{1}{\alpha\cos\theta}\right)+\alpha^{2}\sin^2\theta P(\theta)-\Gamma(\theta)^2\,,\\
0&\, = \, \alpha^2\cos^2\theta \, \Sigma\left(\frac{1}{\alpha\cos\theta}\right)-\Gamma(\theta)\,,
\end{split}
\end{align}
we find 
\begin{align}
\begin{split}
\diff s^2_{3d}\, & = \, -\diff t^2+\diff s^2_{\Sigma} \\
&  =\, -\diff t^2+\frac{\Gamma^4}{P(\Gamma^2-\alpha^2\sin^2\theta\,P)^2}\diff \theta^2+\frac{P\sin^2\theta}{\Gamma^2-\alpha^2 \sin^2 \theta\,P}\diff\phi^2\,.
\end{split}
\end{align}
The two-dimensional metric $\diff s^2_{\Sigma}$ is (a conformal representative of) the metric of the spindle on the conformal boundary,  on a constant time slice.  One can check that $\diff s^2_{\Sigma}$ is singular on the locus defined by
\begin{align}
\Gamma^2\, =\, \alpha^2 \sin^2\theta\,P\,,
\end{align}
which is solved exactly by \eqref{boundary_vs_yA},  namely when the conformal boundary intersects the acceleration horizon.  This can also be argued from the first identity in \eqref{BHidentities}. The boundary spindle then splits into two halves,  with the one for $\theta>\theta_0$ hidden behind the acceleration horizon.  This is precisely the behaviour observed in \cite{Ferrero:2020twa},  where it was shown that this picture can be regulated in various ways,  namely introducing rotation or moving away from supersymmetry and/or extremality.

To conclude this section,  let us discuss the conformal Killing spinors (CKSs) of the boundary metric.  It was noticed in \cite{Ferrero:2020twa} that,  for the Pleba\'nski-Demia\'nski black hole with only mass,  acceleration and magnetic charge,  the CKSs behave differently on the two halves of the boundary,  and in particular they are constant and chiral on one half,  and anti-chiral on the other half,  related to the fact that a topological twist with different relative sign between the spin connection and the gauge field is realized on the two halves of the conformal boundary.  As we shall discuss below,  we observe exactly the same behaviour when two magnetic charges are present.

To show this,  let us introduce the R-symmetry gauge field $A^R=A_1+A_2$,  which reads (in the bulk as well as on the boundary)
\begin{align}
A^R\, = \, \frac{2(1-x^2-\alpha^2)(1-\sqrt{1-\alpha^2}\,\cos\theta)}{\alpha^2\,(1-\alpha^2-x^2\,\cos^2\theta)}\,\diff \phi\,.
\end{align}
We are of course free to make gauge transformations,  and we note that if one defines the equivalent gauge field
\begin{align}\label{AtildeR}
\tilde{A}^R\, = \, A^R-\frac{1-x^2-\alpha^2}{\alpha^2\,(1-\alpha^2)}\,\diff\phi\,,
\end{align}
at the two poles of the spindle (defined by $\theta=0,\pi$),  this satisfies ({\it cf.} the general discussion 
in \cite{Ferrero:2021etw})
\begin{align}
\left.\tilde{A}^R\right|_{\theta=0}\, = \, \frac{1}{n_-}\,\diff\varphi\,,\quad 
\left.\tilde{A}^R\right|_{\theta=\pi}\, = \, \frac{1}{n_+}\,\diff\varphi\,,
\end{align}
where we have introduced $\varphi$ such that $\Delta\varphi=2\pi$,  given by $\varphi=\frac{2\pi}{\Delta\phi}\phi$.  Note in particular that this correctly reproduces the total magnetic flux through the spindle computed in the first line of \eqref{Ps}.  We next introduce a frame
\begin{align}
e^0\, = \, \diff t\,, \quad
e^1\, = \, \frac{\Gamma^2}{\sqrt{P}\,(\Gamma^2-\alpha^2\,P\,\sin^2\theta)}\,\diff\theta\,, \quad
e^2\, = \, \frac{\sqrt{P}\,\sin\theta}{\sqrt{\Gamma^2-\alpha^2\,P\,\sin^2\theta}}\,\diff\phi\,,
\end{align}
and the only non-zero components of the associated spin connection are along the spindle ($\theta,\phi$ directions)
\begin{align}
\omega^{12}_{\Sigma}\, = \, -\frac{(2\cot\theta\,P\,\Gamma+\Gamma\,P'-2P\,\Gamma')\,\sin\theta}{2\,\Gamma\,\sqrt{\Gamma^2-\alpha^2\,P\,\sin^2\theta}}\,\diff\phi\,,
\end{align}
which at the poles satisfies
\begin{align}
\left.\omega^{12}_{\Sigma}\right|_{\theta=0}\, = \, -\frac{1}{n_-}\,\diff\varphi\,, \quad
\left.\omega^{12}_{\Sigma}\right|_{\theta=\pi}\, = \, \frac{1}{n_+}\,\diff\varphi\,, 
\end{align}
in particular giving the correct Euler number \eqref{Euler} for the spindle.  We also note that,  for generic $\theta$,
\begin{align}\label{omega_vs_A}
\omega^{12}_{\Sigma}\, = \, 
\begin{cases}
-\tilde{A}^R\,, \quad (0<\theta<\theta_0)\,,\\
+\tilde{A}^R\,, \quad (\theta_0<\theta<\pi)\,,
\end{cases}
\end{align}
justifying our claim that one half of the conformal boundary realizes a topological twist,  while an anti-topological twist is present on the other half.  To corroborate this claim,  let us also introduce the conformal Killing spinor equation (CKSE)
\begin{align}\label{CKSE}
D_{\mu}\zeta\, = \, \frac{1}{3}\gamma_{\mu}\,\slashed{D}\zeta\,,
\end{align}
where the covariant derivative $D_{\mu}$ contains both the spin connection and the gauge fields\footnote{Here we write $\tilde{A}^R$ for the gauge field to stress that we are using the gauge \eqref{AtildeR}.  Any gauge-equivalent choice is of course equally valid.}
\begin{align}
D\, = \, \diff+\frac{1}{4}\omega^{ab}\,\gamma_{ab}-\frac{\ii}{2}\tilde{A}^R\,,
\end{align}
and we use the gamma matrices
\begin{align}
\gamma_0\, = \, \ii\,\sigma^3\,, \quad
\gamma_1\, = \, \sigma^1\,, \quad
\gamma_2\, = \, \sigma^2\,.
\end{align}
In agreement with the observation \eqref{omega_vs_A} about the relative sign between spin connection and gauge field on the two halves of the conformal boundary,  we find that the solution of \eqref{CKSE} is given by
\begin{align}
\zeta\, = \, \begin{pmatrix}
0\\1
\end{pmatrix}
\quad \text{if} \quad 0<\theta<\theta_0\,, \qquad
\zeta \, = \, 
\begin{pmatrix}
1\\0
\end{pmatrix}
\quad \text{if} \quad
\theta_0<\theta<\pi\,.
\end{align}
Given the chirality matrix $\sigma^3=-\ii\,\gamma_1\gamma_2$,  these are respectively an anti-chiral spinor (in the region where $\omega^{12}_{\Sigma}=-\tilde{A}^R$) and a chiral spinor (in the region where $\omega^{12}_{\Sigma}=+\tilde{A}^R$),  as previously claimed.

\section{Entropy function and BPS relation}\label{sec:conjectures}

Although we only have the full black hole solutions  matching onto the near horizon solutions constructed in section 
\ref{sec:AdS2solutions} in certain special cases, nevertheless in this section we conjecture some general formulas that the 
black holes should satisfy. 
In section~\ref{sec:entropy} we make an educated guess for the on-shell action of the black holes, 
which will allow us to derive again the entropy \eqref{SBH} from the extremization of a suitably defined entropy function. 
Then in section \ref{sec:BPS} we conjecture a BPS formula for the mass of the black holes, again making some non-trivial checks of this 
formula.

\subsection{Entropy function}\label{sec:entropy}

On general grounds, the black hole entropy should be the logarithm of a partition function 
in a microcanonical ensemble, which is then a  function of the conserved charges, as in \eqref{SBH} or \eqref{SBHJ}. 
On the other hand, in AdS/CFT one identifies the holographically renormalized 
on-shell action in gravity with minus the logarithm of the dual field theory
partition function in a grand canonical ensemble. The latter is a function 
of the associated chemical potentials, and the two ensembles are related by a Legendre transform.
While the black hole entropy can be computed from the near horizon 
AdS$_2\times \Sigma$ solutions we have constructed in section \ref{sec:AdS2solutions}, 
in order to compute the on-shell action we in principle need the full (non-extremal) 
black hole solutions,  which in general are not available. 

In this section 
we conjecture a formula for the holographically renormalized on-shell action 
of the full supersymmetric, accelerating, rotating and multi dyonically charged 
black holes, and then make various checks of this conjecture. Firstly, 
we show that it reduces to the correct formula in cases where 
the appropriate families of (complexified) supersymmetric black hole solutions 
are known.  
Secondly, we show that extremizing the associated entropy function, or equivalently 
taking a Legendre transform, we precisely recover the black hole entropy 
\eqref{SBH} we have computed from the near horizon solution. 
In the process we shall also obtain a formula for the angular momentum $J_{BH}$ 
of the black holes, and comment on its relation to $J_{AdS_2}$ given by~\eqref{JAdS2}. 

Our starting point is the following conjecture for the renormalized on-shell action 
of the black holes:
\begin{align}\label{Itwocharge}
I \, = \, I(\omega,\varphi_1,\varphi_2)\, = \,\pm \frac{1}{2\ii  G_{(4)}}\left(16\frac{\varphi_1\varphi_2}{\omega} + \frac{1}{4}P_1P_2\,  \omega\right)\, ,
\end{align}
Here $\omega$ is a rotational chemical potential, while $\varphi_i$ are electric chemical potentials for the 
two gauge fields $A_i$, $i=1,2$. These chemical potentials are furthermore required to satisfy the constraint
\begin{align}\label{constrainttwocharge}
2(\varphi_1 +\varphi_2) -\frac{\chi(\Sigma)}{4}\omega \, = \, \pm \ii \pi\, .
\end{align}
More precisely, \eqref{Itwocharge} should be the holographically renormalized on-shell action 
for a complex locus of supersymmetric solutions, that arise as an analytic continuation 
of the real black hole solutions we have alluded to earlier. We can make the following 
checks of \eqref{Itwocharge} in particular cases:

\begin{enumerate}
\item Recall that the magnetic charges satisfy the constraint
\begin{align}
P_1 + P_2 \, = \, \frac{n_--n_+}{n_-n_+}\, \equiv \, 4 G_{(4)}Q_m\, ,
\end{align}
where $Q_m$ was first introduced in \cite{Ferrero:2020twa}, and we have 
included a factor of $G_{(4)}$ in the definition, as in \cite{Cassani:2021dwa}. Similarly denoting
\begin{align}
Q_1 + Q_2 \, \equiv \, 4 G_{(4)}Q_e\, ,
\end{align}
our near horizon solutions precisely reduce to those studied in \cite{Ferrero:2020twa} on 
setting
\begin{align}
P_1\, = \, P_2 \, = \, 2G_{(4)}Q_m\, , \qquad Q_1 \, = \, Q_2\, = \, 2G_{(4)}Q_e\, .
\end{align}
The thermodynamics and on-shell action of the associated full black hole 
solutions were studied recently in \cite{Cassani:2021dwa}, and 
correspondingly setting $\varphi_1=\varphi_2=\varphi/4$, 
the formula \eqref{Itwocharge} reduces to that derived in this reference, 
as does the constraint~\eqref{constrainttwocharge}.
\item Instead setting $P_1=P_2=0$, and formally setting $n_-=n_+=1$ so that 
the spindle becomes $\Sigma=S^2$, we should recover the Kerr-Newman 
family of electrically charged rotating black holes studied in \cite{Cassani:2019mms}. 
Comparing to the latter reference, we write
\begin{align}
Q_1 \, = \, 8Q_1^{\mathrm{there}}\, , \quad Q_2 \, = \, 8Q_3^{\mathrm{there}}\, , \quad 
\varphi_1\, = \, \frac{1}{4}\varphi_1^{\mathrm{there}}\, , \quad \varphi_2\, = \, \frac{1}{4}\varphi_3^{\mathrm{there}}\, .
\end{align}
Again, our conjectured on-shell action \eqref{Itwocharge} and constraint then indeed reduce to those derived 
in \cite{Cassani:2019mms} for the full black hole solutions.
\item Finally, setting $P_1=-P_2=P$, so that $Q_m=0$ and formally $n_-=n_+=1$ so that again the spindle 
becomes $\Sigma=S^2$, the corresponding entropy function we write down 
in \eqref{Stwocharge} below agrees with that  proposed  in \cite{Hristov:2019mqp,Hosseini:2020mut} 
for a family of Kerr-Newman black holes with two electric charges $Q_1,Q_2$, 
and a magnetic charge variable $P$.  The same solution is also discussed as a subcase of our general solution \eqref{spinningspindle}, \eqref{spinningspindlefunctions} in appendix \ref{app:dyonicnonacc}.  We note that in referencs  \cite{Hristov:2019mqp,Hosseini:2020mut}  this 
entropy function was also conjectured, rather than derived 
directly from a renormalized action. One can think 
of the solutions discussed in this paper as generalizing those 
in \cite{Hristov:2019mqp,Hosseini:2020mut}   by adding acceleration and 
an anti-topological twist magnetic flux $Q_m\neq 0$.
\end{enumerate}

Given the on-shell action \eqref{Itwocharge}, we may write down the following 
associated \emph{entropy function}
\begin{align}\label{Stwocharge}
\mathcal{S} \equiv \,  -  I(\omega,\varphi_1,\varphi_2) - \frac{1}{G_{(4)}}\left(\omega J_{BH} + \varphi_1\,  Q_1+\varphi_2\, Q_2\right)\, .
\end{align}
Here the rotational chemical potential $\omega$ is conjugate to the black hole angular momentum 
$J_{BH}$. According to the discussion at the start of this subsection, the black hole entropy should then be obtained by extremizing 
\eqref{Stwocharge} over the chemical potentials $\omega$, $\varphi_i$, where the latter are 
subject to the constraint \eqref{constrainttwocharge}. This of course then implements the 
Legendre transform. We thus write 
\begin{align}\label{entropyext}
S(J_{BH},Q_1,Q_2)\, = \, \mathrm{ext}_{\{\omega,\varphi_1,\varphi_2,\Lambda\}}
\left[ \mathcal{S} - \Lambda\left(2(\varphi_1+\varphi_2)-\frac{\chi(\Sigma)}{4}\omega \mp \ii \pi\right)\right]\, .
\end{align}
The extremization imposes
\begin{align}
-\frac{\partial I}{\partial\omega} \, = \, \frac{J_{BH}}{G_{(4)}} - \frac{\chi(\Sigma)}{4}\Lambda\, , \qquad 
-\frac{\partial I}{\partial\varphi_i}\, = \, \frac{Q_i}{G_{(4)}} + 2\Lambda\, ,
\end{align}
and we find the solution
\begin{align}
\begin{split}
\Lambda \, & = \, \frac{1}{4G_{(4)}}\Big\{-Q_1-Q_2 \pm \ii \chi(\Sigma) + \ii \eta \Big[\chi(\Sigma)^2-(Q_1-Q_2)^2+4P_1P_2 \\ & \qquad \qquad \pm 2\ii (Q_1+Q_2)\chi(\Sigma)\pm 32\ii\,  J_{BH}\Big]^{1/2}\Big\}\, , \\
\omega\, & = \, \frac{4\pi \ii \eta}{\sqrt{\chi(\Sigma)^2-(Q_1-Q_2)^2+4P_1P_2 \pm 2\ii (Q_1+Q_2)\chi(\Sigma)\pm 32\ii\,  J_{BH}}}\, , \\
\varphi_1\, & = \, \pm \frac{\pi \eta[ Q_2-Q_1\pm \ii \chi(\Sigma)] }{4\sqrt{\chi(\Sigma)^2-(Q_1-Q_2)^2+4P_1P_2 \pm 2\ii (Q_1+Q_2)\chi(\Sigma)\pm 32\ii\,  J_{BH}}}\pm \frac{\ii\pi}{4}\, .
\end{split}
\end{align}
Here $\varphi_2$ is determined by the constraint \eqref{constrainttwocharge}, and $\eta=\pm 1$ arises 
as a choice of sign in taking square roots when solving the equations. 
Imposing that the entropy is real, while assuming all conserved charges are real,  we find 
\begin{align}\label{JBH}
  J_{BH} &\, = \,  \frac{Q_1+Q_2}{16}\left[-\chi(\Sigma) + \sqrt{\chi(\Sigma)^2+4(P_1P_2+Q_1Q_2)}\right]\, ,
   \end{align}
where the sign of the square root in $J_{BH}$ has been fixed by requiring $J_{BH}>0$. 
Moreover, we then obtain the extremal value of \eqref{entropyext} to be  
\begin{align}
S(J_{BH},Q_1,Q_2) \, & = \,  \pm \ii \pi \Lambda\, ,
\end{align}
which gives
\begin{align}
\begin{split}
S(J_{BH},Q_1,Q_2) \, & = \,  \frac{4\pi} {G_{(4)}(Q_1+Q_2)} J_{BH} \\
& = \, \frac{\pi}{4G_{(4)}}\left[-\chi(\Sigma) + \sqrt{\chi(\Sigma)^2+4(P_1P_2+Q_1Q_2)}\right] \, .\label{babel}
\end{split}
\end{align}
This precisely agrees with the black hole entropy $S_{BH}$ in \eqref{SBH} computed from the near horizon solutions.

Notice that the two angular momenta \eqref{JAdS2}, \eqref{JBH} are related via
\begin{align}\label{JorJ}
J_{AdS_2} - \frac{4Q_1Q_2}{(Q_1+Q_2)^2}J_{BH} \, = \, \frac{Q_1Q_2}{4(Q_1+Q_2)}\chi(\Sigma)\, .
\end{align}
Since we do not have the full black hole solutions it is not immediate to define and compute 
$J_{BH}$ directly. However, we note that both $J_{AdS_2}$ and $J_{BH}$ were 
computed for the minimal gauged supergravity solutions with $Q_1=Q_2$ (and $P_1=P_2$) in 
\cite{Ferrero:2020twa}, and the relation \eqref{JorJ} reduces to the corresponding relation
in this reference. Moreover, the analogous quantities can be computed 
for the Kerr-Newman black holes in \cite{Cassani:2019mms}, and we find 
\begin{align}
J_{AdS_2} - \frac{4Q_1Q_2}{(Q_1+Q_2)^2}J_{BH} \, = \, \frac{Q_1Q_2}{2(Q_1+Q_2)} \qquad \mbox{(Kerr-Newman)}\, .
\end{align}
This is precisely equation \eqref{JorJ}, with $\chi(\Sigma)=\chi(S^2)=2$, and is thus another 
consistency check on this formula.

We end this subsection making two additional comments. We note that there exists a purely accelerating (non-rotating) dyonic configuration with $Q_1+Q_2=0$, mirroring the
purely rotating (non-accelerating) dyonic configuration with $P_1+P_2=0$, discussed in the third item above.\footnote{These two families of solutions intersect for $Q_1+Q_2=0=P_1+P_2$, corresponding to a dyonic static black hole, with horizon a non-rotating two-sphere.}  Despite the fact that in this case 
the angular momentum vanishes, 
the chemical potential $\omega$ is non-zero, and the entropy function
takes simply the form 
\eqref{Stwocharge}, with  $Q_1=- Q_2 = Q$.  Extremizing this leads to the entropy 
\begin{align}
\label{gluingshit}
\begin{split}
S(J_{BH},Q) \, & =  \, \frac{\pi}{4G_{(4)}}\left[-\chi(\Sigma) + \sqrt{\chi(\Sigma)^2+4(P_1P_2-Q^2)}\right] \, ,
\end{split}
\end{align}
that is consistent with the second equality in \eqref{babel}.

We also note that our proposed entropy function can be expressed as the Legendre transform of  
\begin{align}
\label{speculateonI}
I(\omega,\varphi_1,\varphi_2) \, = \, \pm \frac{2}{\ii \pi \, \omega} \left[ F_{S^3}(\varphi_i - \tfrac{1}{8}\omega P_i) +  F_{S^3}(\varphi_i + \tfrac{1}{8}\omega P_i)  \right]\, , 
\end{align}
where $F_{S^3}(\Delta_1,\Delta_2)=4 \Delta_1 \Delta_2F_{S^3}$ is the 
large $N$  $S^3$  free energy as a function of the trial R-symmetry, with $\Delta_i$ satisfying $\Delta_1+\Delta_2=1$. 
Recall here that the free energy on the three-sphere is  $F_{S^3}=\frac{\pi}{2G_{(4)}}=\frac{\sqrt{2}\pi N^{3/2}}{3}$. 
This is consistent with the general expectations for  $d=3$,  ${\cal  N} =4$ SCFT with holographic duals, discussed in \cite{Hosseini:2020mut}. 
It is tempting to speculate that for ${\cal N}=2$ SCFTs, with four different chemical potentials $\Delta_i$ subject to  $\Delta_1+\Delta_2+\Delta_3+\Delta_4=2$, the expression \eqref{speculateonI} remains valid, and the entropy function is the obvious 
four-charge extension of \eqref{Stwocharge}, with constraints
 \begin{align}
  \label{moreofit}
  \begin{split}
\sum_{i=1}^4 P_i  &\, = \,\frac{n_--n_+}{n_-n_+}\,\, , \\
 \sum_{i=1}^4\varphi_i  - \frac{\chi (\Sigma)}{4}\omega & \, = \,  \pm  \ii\pi\, ,
\end{split}
\end{align}
where recall that for the ABJM model  $F_{S^3}(\Delta_1,\ldots,\Delta_4)=4\sqrt{\Delta_1\Delta_2\Delta_3\Delta_4}\, F_{S^3}$. 
This gives a prediction for the entropy and charges of a conjectural dyonic, rotating and accelerating,  four-charge solution of STU supergravity. 
We also note that the conjectured formula \eqref{speculateonI} for the on-shell action is very suggestive 
of a localization formula, with the two terms arising from contributions at the two poles of the spindle 
horizon of the complexified black hole solution. For the solution in minimal gauged supergravity, in which 
the magnetic charges and chemical potentials are all equal, this is precisely the case -- see equation (5.20) of \cite{Cassani:2021dwa}, 
which uses the localization formula of \cite{BenettiGenolini:2019jdz}. 

\subsection{BPS relation}\label{sec:BPS}

The standard holographic approach to defining the mass of black holes in AdS, as well as other 
conserved charges, involves first computing the boundary holographic energy-momentum tensor and 
conserved currents, the latter being associated with the global $U(1)$ symmetries dual
to the gauge fields $A_i$ in the bulk, $i=1,2$. The mass is then defined as the conserved 
charge associated with a time-like Killing vector for the solution. This was the approach 
taken in \cite{Cassani:2021dwa} for the minimal gauged supergravity black holes with $Q_1=Q_2$, $P_1=P_2$. 
However, this definition of mass then involves a choice of time-like Killing vector, which is not unique. 
This ambiguity was fixed in \cite{Cassani:2021dwa} by requiring the first law of black hole thermodynamics to hold, which 
effectively fixed the choice of time-like Killing vector. 
Since we do not have the general black hole solutions, clearly we cannot follow this approach here. 
One might hope that, as with the angular momentum, one could alternatively define 
the mass using an appropriate Komar integral, which moreover could be evaluated 
on the horizon, rather than at the conformal boundary. Instead in this subsection we shall simply 
conjecture a BPS formula for the mass of the black holes, and leave these  interesting questions 
about how to define and compute the mass more generally for future work.

Our conjectured BPS formula is
\begin{align}\label{Mmin}
M \, = \, \frac{2}{\chi}J_{BH} + \frac{1}{4}(Q_1+Q_2)\, ,
\end{align}
where the $J_{BH}$ is given by \eqref{JBH}. 
This reduces to the BPS formulas for the masses of the black holes 
studied in both \cite{Cassani:2021dwa} and \cite{Cassani:2019mms}. Substituting in for $J_{BH}$ using \eqref{JBH}, or 
$J_{AdS_2}$ using \eqref{JAdS2}, we obtain 
\begin{align}
\begin{split}
M \, & = \, \frac{\chi +  \sqrt{\chi^2+4(P_1P_2+Q_1Q_2)}}{8\chi} (Q_1+Q_2)\\
& = \, \frac{2}{\chi}\frac{(Q_1+Q_2)^2}{4Q_1Q_2}J_{AdS_2} + \frac{1}{8}(Q_1+Q_2)\, .
\end{split}
\end{align}

\section{Discussion}\label{sec:discussion}

The Pleba\'nski-Demia\'nski  solutions \cite{Plebanski:1976gy} describe the most general dyonically charged, rotating and accelerating 
black holes \cite{Podolsky:2006px} in four-dimensional Einstein-Maxwell theory, or equivalently minimal 
$D=4$, $\mathcal{N}=2$ gauged supergravity. However, beyond these solutions relatively little is known 
about accelerating black holes in other theories, or accelerating black objects more generally. Nevertheless, it is clear that such solutions exhibit interesting 
properties, including unconventional horizon topologies, and extended thermodynamics (see \cite{Cassani:2021dwa, Anabalon:2018qfv} and  references therein). For asymptotically locally AdS solutions the conical deficit singularities of the horizon extend 
out to the conformal boundary, and the dual field theory is then defined on a (mildly) singular background. 
This leads to the interesting question of how to define and study (supersymmetric) field theories on such 
backgrounds. 
Similarly, for extremal solutions the near horizon limits are solutions of the type AdS$_2\times \Sigma$, where $\Sigma$
 also have conical deficit/orbifold singularities.
In this paper we have constructed a very general class of rotating AdS$_2\times \Sigma$ 
solutions to $D=4$, $\mathcal{N}=4$ gauged supergravity, and have argued that these 
should arise as near horizon limits of a general family of multi-dyonically charged, rotating and accelerating
black holes. Although in this paper we have constructed new (supersymmetric or otherwise)
accelerating black hole solutions to this theory, carrying magnetic charge, we have so far been 
unable to  construct the general family, carrying also electric charges and angular momentum. 
It remains an important outstanding problem to find these black hole solutions, or at least show they exist. 
The near horizon rotating AdS$_2\times \Sigma$ solutions of section \ref{sec:AdS2solutions}, once uplifted on $S^7$ to 
$D=11$ supergravity, should fit into the recent classification of  \cite{Couzens:2020jgx}, 
and it would be interesting to check this explicitly.

The results of this paper generalize those of \cite{Ferrero:2020twa} to non-minimal $D=4$ gauged supergravity, 
in particular to theories with multiple gauge fields and hence multiple conserved electric and magnetic charges. 
In contrast to the Pleba\'nski-Demia\'nski  solutions of the minimal theory, here there exist (near horizon)
supersymmetric and extremal \emph{dyonic} solutions that are either  rotating, or accelerating. 
Our results share some features with those in references \cite{Hosseini:2021fge, Boido:2021szx},   which 
generalize the original AdS$_3\times \Sigma$ D3-brane spindle solutions of minimal $D=5$ gauged supergravity 
in \cite{Ferrero:2020laf} to STU supergravity. In particular, we have found that the total magnetic flux 
is given by a so-called anti-topological twist, with the Killing spinors being 
sections of non-trivial bundles over the horizon. 
In appendix~\ref{app:fourcharges} we have presented the local form of (non-rotating) 
accelerating  black holes and of a  family  of supersymmetric
AdS$_2\times \Sigma$ spindle solutions with four magnetic charges\footnote{Once uplifted to $D=11$, these solutions correspond to those previously found in \cite{Gauntlett:2006ns}.}, that are solutions to STU gauged supergravity.  
We believe that, as in the two-charge sub-family, the latter solutions should arise as the near horizon limit of the accelerating black holes, in the extremal limit. 
However, after a preliminary analysis, unlike in \cite{Boido:2021szx,Hosseini:2021fge} we have been unable to solve all of the regularity conditions in closed form, in either set of solutions. 
 We therefore leave for the future 
 a full analysis of the global regularity conditions 
of these solutions and the proof that the four-charge AdS$_2\times \Sigma$  spindle solutions  arise in the near-horizon limit of  the supersymmetric  black~holes. 

We recall that
the $D=5$ AdS$_3\times\Sigma$ solutions  of 
\cite{Ferrero:2020laf}
 have no known corresponding 
 supersymmetric and extremal  accelerating black string solutions, of which they are near horizon limits, 
 and finding such solutions is again an interesting open problem.

Building on the results of \cite{Cassani:2021dwa} for the solutions in the minimal theory, we have presented a conjectural entropy function, that reproduces the entropy of the (conjectural) black holes after extremizing it and imposing reality conditions. 
Lacking the explicit non-extremal black hole solutions, we made several assumptions that would clearly be desirable to corroborate with further work. First of all, the 
entropy function \eqref{Stwocharge} makes use of the expression 
\eqref{Itwocharge}, which is the conjectured on-shell action of the corresponding class 
of supersymmetric but non-extremal (complexified) black hole solutions.
We expect that it should be possible to bypass the need for the explicit non-extremal solutions, by extending the holographic localization results of \cite{BenettiGenolini:2019jdz} to non-minimal theories. 
Related to this, we have arrived at expressions for the angular momentum and mass of the black holes with some educated guesses, which pass various consistency checks. However, it would be nice to be able to extract directly from the supersymmetric near horizon solutions the 
correct mass and angular momentum, in analogy with the electric and magnetic charges. In particular, we suspect that 
the  ``AdS angular momentum'' introduced in  \cite{Cassani:2021dwa} should have a direct interpretation in the holographically dual field theory, which would be interesting to investigate further.

As we discussed in section \ref{sec:entropy}, the entropy function that we proposed reduces to various notable cases for special values of the parameters. Differently from the minimal setup, in the multi-charge setting we can have supersymmetric dyonic configurations, which are either only rotating or only accelerating. In the purely rotating case, the horizon is necessarily spherical and the entropy function can be naturally interpreted in terms of two electric chemical potentials and the angular velocity $\omega$, subject to the constraint \eqref{constrainttwocharge}
\cite{Hosseini:2020mut}. In the purely accelerating case, the horizon is a spindle
and the angular momentum is zero. Nevertheless, we find that the chemical potential $\omega$ is non-zero for these solutions, and in order to reproduce the entropy,
 the entropy function keeps the form \eqref{Itwocharge}, where one uses $Q_1+Q_2=0$. 

Finally, it would be interesting to analyse the field theory duals of the solutions we have discussed in this paper. 
  The starting point of this analysis should be a ``flavoured'' version of the supersymmetric
  partition function anticipated in  \cite{Cassani:2021dwa}, arising from an (anti-)twisted compactification of the $d=3$ theories on a spindle, that can be denoted as 
  $Z(n_+,n_-,\varphi_i,\omega)_{S^1\times\Sigma}$.   In \eqref{speculateonI} (subject to \eqref{moreofit}) we wrote our prediction for the large $N$ limit of $-\log  Z(n_+,n_-,\varphi_i,\omega)_{S^1\times\Sigma}$, for an arbitrary 
  ${\cal N}=2$, $d=3$ SCFT with a Freund-Rubin dual, compactified on a spindle.
 
\subsection*{Acknowledgments}
We would like to thank Minwoo Suh for pointing out some typographical errors in an earlier version of this manuscript. This work was supported by STFC grant  ST/T000864/1.

\appendix

\section{AdS$_2$ solutions from an ansatz}\label{app:ansatz}

In this appendix we give more details about the ansatz that leads to the supersymmetric solutions described in section \ref{sec:AdS2solutions}. In the following subsections we then describe some cases that arise as special limits of the general solution.

\subsection{The ansatz}\label{app:ansatzsolution}

We are looking for solutions of the $D=4$, $\mathcal{N}=4$ supergravity model described in section \ref{sec:AdS2solutions} of the form AdS$_2 \times \Sigma$  that are invariant under the isometries of AdS$_2$,  where $\Sigma$ is a two-dimensional Riemannian space with a $U(1)$ isometry.  One can show that the most general metric and gauge fields satisfying the above requirements are given precisely by those in \eqref{spinningspindle},  namely
\begin{align}\label{spinningspindleappendix}
\begin{split}
\diff s^2_4&\, = \, \frac{1}{4}\lambda(y)\,\left(-\rho^2\,\diff\tau^2+\frac{\diff\rho^2}{\rho^2}\right)+\frac{\lambda(y)}{q(y)}\,\diff y^2+\frac{q(y)}{4\,\lambda(y)}\,(\diff z+\jp\,\rho\,\diff \tau)^2\,,\\
A_i&\, = \, \frac{h_i(y)}{\lambda(y)}\,(\diff z+\jp\,\rho\,\diff \tau)\,, \quad
\ex^{\xi}\, = \, \frac{g_1(y)}{\lambda(y)}\,, \quad
\chi\, = \, \frac{g_2(y)}{g_1(y)}\,,
\end{split}
\end{align}
where at this stage all the functions should be thought of as unknowns.  Demanding that the equations of motion are satisfied leads to non-linear ordinary differential equations for the functions $\lambda(y)$,  $q(y)$,  $h_i(y)$ and $g_i(y)$ ($i=1,2$),  of which we were not able to find the general solution.  Inspired by the structure of some known black hole solutions,  we then make the ansatz that all the functions appearing explicitly in \eqref{spinningspindle} are polynomials in $y$,  whose degree is also fixed in analogy with other solutions.  More precisely,  we set
\begin{align}
\begin{split}
\lambda(y)&\, = \, y^2+\Delta\,, \qquad
h_2(y)\, = \, \sum_{n=0}^2 b_n y^n\,,\quad
h_1(y)\,  =\, \sum_{n=0}^2 a_n y^n\,,\\
q(y)&\, = \, \sum_{n=0}^4x_n y^n\,,\quad \hspace{0.1cm}
g_1(y)\, = \, \sum_{n=0}^2 \alpha_n y^n\,,\quad \hspace{-0.05cm}
g_2(y)\, = \, \sum_{n=0}^2 \beta_n y^n\,,
\end{split}
\end{align}
and we find that the equations of motion constrain the constants introduced above in such a way that the solution only depends on six parameters.  Note that we are not yet requiring supersymmetry at this stage.  The explicit solution reads 
\begin{align}
\begin{split}
\lambda(y)&\, = \, y^2+\jp^2-\tfrac{1}{4}(\alpha_1^2+\beta_1^2)\,,\\
q(y)&\, = \, (y^2+\jp^2)^2+\tfrac{1}{2}(8\jp^2-8-\alpha_1^2-\beta_1^2)y^2\\
&+\frac{4 \left[\left(a_1^2-b_1^2-4 \jp^2 \left(a_2^2-b_2^2\right)\right)\alpha _1 +4 \jp \left(a_1  a_2-b_1 b_2\right) \beta_1\right]}{\alpha _1^2+\beta _1^2} y+\frac{ \left(\alpha _1^2+\beta _1^2\right)^2}{16}\\
&-8 \jp^2 \left(a_2^2+b_2^2\right)-2 \left(a_1^2+b_1^2\right)+\frac{\alpha _1^2+\beta _1^2}{2} \left(2-3 \jp^2\right) +4 \left(1-\jp^2\right) \jp^2\,,\\
h_1(y)&\, = \, a_2 \lambda(y)+a_1 y-\frac{a_1 \alpha_1}{2}-\jp (2 \jp+\beta_1) a_2\,,\\
h_2(y)&\, = \, b_2 \lambda(y)+b_1 y+\frac{b_1\alpha_1}{2}-\jp\,(2\jp-\beta_1)b_2\, ,\\
g_1(y)&\, = \, y^2+\alpha_1 y+\frac{\alpha_1^2+\beta_1^2}{4}+\jp (\jp-\beta_1)\,,\\
g_2(y)&\, = \, \beta_1 y+\jp \alpha_1\,,
\end{split}
\end{align}
where we have used seven parameters ($a_{1,2}$,  $b_{1,2}$, $\alpha_{1}$, $\beta_{1}$ and $\jp$),  which are actually subject to the constraint
\begin{align}\label{constraint}
\begin{split}
0\, & =\, 4 \alpha _1 \jp \left(a_1 a_2-b_1 b_2\right)-\left(a_1^2-b_1^2\right)\beta _1+4\jp^2 \left(a_2^2-b_2^2\right)\beta _1\\ 
& \quad\  +2 \jp \left(1-\jp^2\right)\, \left(\alpha _1^2+\beta _1^2\right)\,.
\end{split}
\end{align}
One can think of this solution as the near horizon limit of an extremal but non-supersymmetric black hole,  that has seven parameters (mass, acceleration, angular momentum and two pairs of dyonic charges) constrained by the extremality condition.

We now demand that this solution to the equations of motion also preserves supersymmetry.  In practice,  this can be done by writing the gaugino Killing spinor equation as
\begin{align}
\delta \lambda\, =\, 0\, = \, \mathcal{M}\,\epsilon\,,
\end{align}
where $\epsilon$ is a Killing spinor and the matrix $\mathcal{M}$ is implicitly determined from \eqref{KSEs}.  A necessary condition for a solution to preserve some supersymmetry is that
\begin{align}
\det \mathcal{M}\,=\,0\,,
\end{align}
which together with the constraint \eqref{constraint} fixes
\begin{align}
b_1\, = \, a_1+\sqrt{1-\jp^2}\alpha_1\,, \quad
b_2\, = \, \sqrt{1-\jp^2}-a_2\,,\quad
\beta_1\, = \, 2\jp-\frac{4a_2}{\sqrt{1-\jp^2}}\jp\,.
\end{align}
The latter conditions turn out to be also sufficient for supersymmetry to be preserved, and indeed the explicit Killing spinors can be found in \eqref{M2spinors}. This leaves us with a supersymmetric solution that depends on the four parameters $a_{1,2}$,  $\alpha_1$ and $\jp$.  Finally,  we can match the parametrization used in \eqref{spinningspindlefunctions} simply using 
\begin{align}
\begin{split}
a_1\, & = \, -\frac{c_1}{2}-\sqrt{1-\jp^2}\sqrt{2c_2-c_3^2\jp^2}\,, \qquad a_2\,  = \, \frac{1-c_3}{2}\sqrt{1-\jp^2}\,,\\
\alpha_1\, & = \, 2\sqrt{2c_2-c_3^2\jp^2}\,.
\end{split}
\end{align}

\subsection{Dyonic,  non-accelerating case}\label{app:dyonicnonacc}

In minimal gauged supergravity the BPS conditions couple the electric charge to rotation and the magnetic charge to acceleration: if one parameter of the two pairs is turned off,  the other is automatically set to zero by the supersymmetry conditions.  This is however not true when more than one gauge field is present,  as we shall now discuss.  In particular,  in this subsection we shall focus on the case of vanishing acceleration,  showing that supersymmetry still allows for non-vanishing magnetic charges.

While we do not have an explicit expression for the acceleration parameter,  it is known that this is responsible for the presence of conical deficits.  Hence,  the natural way to turn off acceleration is to set $n_{\pm}=1$ in the regularity equations.  Note that in this case the parameter $c_1=0$ (see \eqref{quantiseparameters}),  and most importantly from the first equation in \eqref{Ps} the two magnetic charges satisfy
\begin{align}
P_1+P_2=0\,.
\end{align}
However,  from the second equation in \eqref{Ps} we see that the difference between the magnetic charges is independent of $c_1$.  Hence,  we conclude that when two charges are present supersymmetry allows for non-accelerating dyonic black holes,  provided that the total magnetic flux vanishes.  This is precisely the case considered in \cite{Hristov:2019mqp},  and indeed in this case our entropy \eqref{SBH} matches that in equation  (54) of \cite{Hristov:2019mqp},  after setting $P_1=-P_2=P^{\mathrm{there}}$,  $\chi(\Sigma)=2$ and adjusting the normalizations.  Moreover,  this comparison also constitutes further evidence supporting our claim that $J_{BH}$ in \eqref{JBH} is the correct angular momentum of the black hole associated with our general solution,  as it correctly reduces to $\mathcal{J}^{X^0\,X^1}$ of \cite{Hristov:2019mqp} after the same identifications that are required to match the entropy.

\subsection{Dyonic,  non-rotating case}\label{app:dyonicnonrot}

Inspired by the observations of the previous subsection,  one could consider a mirrored case in which the rotation parameter is turned off.  Naively,  equation \eqref{Qs} tells us that in the non-rotating limit $\mathtt{j}=0$ the sum of the electric charges vanishes,  and since the difference is proportional to the sum then the difference has to vanish as well.  We would thus conclude that there are no dyonic black holes in the non-rotating case.

However,  this is not correct,  as we discuss below.
 If we introduce a new parameter $c_4\equiv c_3\,\mathtt{j}$,  and only then take $\mathtt{j}\to 0$,  we can see that this sets 
\begin{align}
Q_1+Q_2=0\,, \qquad
Q_1-Q_2=-2\,c_4\,\frac{\Delta z}{2\pi}\,,
\end{align} 
so that the difference between the electric charges can indeed be non-vanishing.  At the level of the solution \eqref{spinningspindle},  if we take $c_4=c_3\,\mathtt{j}$ and then send $\mathtt{j}\to 0$ we obtain
\begin{align}\label{nonspinningspindle}
\begin{split}
\diff s^2_4&\, = \, \frac{1}{4}\lambda(y)\,\left(-\rho^2\,\diff\tau^2+\frac{\diff\rho^2}{\rho^2}\right)+\frac{\lambda(y)}{q(y)}\,\diff y^2+\frac{q(y)}{4\,\lambda(y)}\,\diff z^2\,,\\
A_i&\, = \, \frac{h_i(y)}{\lambda(y)}\,\diff z -\eta_i\,\frac{c_4}{2}\,\rho\,\diff \tau\,, \quad
\ex^{\xi}\, = \, \frac{g_1(y)}{\lambda(y)}\,, \quad
\chi\, = \, \frac{g_2(y)}{g_1(y)}\,,
\end{split}
\end{align}
where $\eta_1=+1$ and $\eta_2=-1$,  with the functions given by
\begin{align}\label{spinningspindlefunctionsdyonicnorot}
\begin{split}
\lambda(y)\,= \, &  \, y^2-2c_2\,,\\
q(y)\,=\,& \, y^4-4(1+c_2)\,y^2+4\,c_1\,y+4\,c_2^2-c_1^2\,,\\
h_1(y)\,=\,&\,\frac{1}{2}\left(y^2-(c_1+2\,\sqrt{2\,c_2-c_4^2})\,y+2\,c_2+\sqrt{2\,c_2-c_4^2}\,c_1\right)\,,\\
h_2(y)\,=\,&\,\frac{1}{2}\left(y^2-(c_1-2\,\sqrt{2\,c_2-c_4^2})\,y+2\,c_2-\sqrt{2\,c_2-c_4^2}\,c_1\right)\,,\\
g_1(y)\,=\,&\, y^2+2\,\sqrt{2\,c_2-c_4^2}\,y+2\,c_2\,,\\
g_2(y)\,=\,&\, 2\,c_4\,y\,.
\end{split}
\end{align}
Note that if one simply takes $\mathtt{j}\to 0$ the two gauge fields appear to develop a singularity.  However,  this can easily be seen to be due to a pure gauge term,  and it can be cured by adding to $A_i$ in \eqref{spinningspindle} a term $\eta_i\,\frac{c_4}{2\,\mathtt{j}}\diff z$.  It is straightforward to check that this is a solution of the equations of motion,  and it preserves the same amount of supersymmetry as \eqref{spinningspindle},  with the Killing spinors obtained from \eqref{M2spinors} after setting $c_4=c_3\,\mathtt{j}$ and sending $\mathtt{j}\to 0$.  The entropy of this solution is  obtained from the general formula \eqref{SBH},  after setting $Q_1=-Q_2$.

\subsection{Purely magnetic case}\label{app:purelymag}

As a special case of the general solution \eqref{spinningspindle},  we can consider the limit in which the black hole has only magnetic charges,  with no rotation.  In this case,  local AdS$_2\times \Sigma$ solutions can be extracted from the AdS$_2\times Y_9$ solutions of $D=11$ supergravity discussed in section 5.4 of \cite{Gauntlett:2006ns}.  Using the uplift formulas of \cite{Azizi:2016noi},  one can interpret the latter as solutions of the $D=4$ gauged supergravity,  given by
\begin{align}\label{2chargemagneticAdS2}
\begin{split}
\diff s^2_4&\, = \, \frac{1}{4}\Lambda(w)\left(-\rho^2\diff\tau^2+\frac{\diff\rho^2}{\rho^2}\right)+\frac{\Lambda(w)}{Q(w)}\diff w^2+\frac{Q(w)}{4\Lambda(w)}\diff \z^2\,,\\
A_i&\, =\, \frac{w}{w+q_i}\diff z\,, \quad \ex^{\xi}\, = \, \frac{w+q_1}{w+q_2}\,, \quad \chi\, = \, 0\,,
\end{split}
\end{align}
where
\begin{align}
\Lambda(w)\, = \, (w+q_1)(w+q_2)\,, \qquad
Q(w)\, = \, \Lambda(w)^2-4w^2\,.
\end{align}
The structure of this solution is clearly analogous to that of \eqref{spinningspindle}: the two are indeed diffeomorphic provided that we set
\begin{align}
y \, =  \, w+\frac{q_1+q_2}{2}\,, \quad
c_1\, = \, q_1+q_2\,, \quad
c_2\, = \, \frac{(q_1-q_2)^2}{8}\,,  \quad \jp\, = \, 0\,,
\end{align}
where the gauge fields are identified only up to a gauge transformation,  and the parameter $c_3$ becomes unphysical in this limit,  as it appears only in a pure gauge term for the one-forms $A_i$.

Finally,  let us remark that this case with only magnetic charge and acceleration corresponds to the near horizon limit of the supersymmetric black holes presented in section \ref{sec:BH},  as  will be proved in appendix \ref{app:NHlimit}.

\subsection{Minimal gauged supergravity}\label{app:minimal}

Here we briefly consider the limit in which our solution is a solution of minimal gauged supergravity,  namely when the scalars $\xi$ and $\chi$ vanish,  while the two gauge fields are equal.  Looking at the solution \eqref{spinningspindle} and the functions \eqref{spinningspindlefunctions},  one can see that the choice $c_2=c_3=0$ sets to zero the scalars and gives $A_1=A_2$.  It is then straightforward to see that our multi-charge spindle solution reproduces in this limit the AdS$_2$ solutions discussed in \cite{Ferrero:2020twa},  if one sets $c_1=\mathtt{a}$,  with all the other coordinates and parameters unchanged.

\section{Near horizon limit of accelerating black holes}\label{app:NHlimit}

In this section we show that,  in the supersymmetric and extremal case,  the near horizon limit of the black hole solution \eqref{magneticBH} is a special case of the class of AdS$_2\times \Sigma$ solutions discussed in section \ref{sec:AdS2solutions},  namely the case $c_3=0=\jp$,  in which the electric charges and the rotation parameter are turned off. Note that the near-horizon solution in this case was also discussed in appendix \ref{app:purelymag}.

To prove this,  we start from the solution \eqref{magneticBH} and the functions \eqref{functionsmagneticBH},  which is supersymmetric when the parameters satisfy \eqref{magneticBHsusy}.  Note that the supersymmetry conditions also imply extremality.  To find the near horizon metric,  we change coordinates using
\begin{align}\label{toNH}
r \, \to  \, r_+ + \lambda s \rho\,,  \qquad
t\,  \to \, \lambda^{-1} s \tau\,,
\end{align}
where $s$ is a constant,  and then we take the limit $\lambda\to 0$.  We normalize the AdS$_2$ metric choosing 
\begin{align}
s\, = \, \frac{\sqrt{\xp^2-\alpha^2(1-\alpha^2)r_+^2}}{(r_+-r_-)\alpha(1-\alpha^2)}\,,
\end{align}
and we find the near horizon metric
\begin{align}\label{AdS2metricfromBH}
\begin{split}
\diff s^2\, = \, &\, \frac{(\xp^2-\alpha^2(1-\alpha^2)r_+^2)(1-\alpha^2-\xp^2\cos^2\theta)}{(r_--r_+)^2\alpha^2(1-\alpha^2)^3(1-\alpha r_+ \cos\theta)^2}\left(-\rho^2\diff \tau^2+\frac{\diff \rho^2}{\rho^2}\right)\\
&+\frac{r_+^2-\frac{\xp^2}{\alpha^2(1-\alpha^2)}}{(1-\alpha r_+ \cos\theta)^2}\left(\frac{\Gamma}{P}\diff\theta^2+\frac{P}{\Gamma}\sin^2\theta \diff\phi^2\right)\,,
\end{split}
\end{align}
where the functions $P$ and $\Gamma$ are those of \eqref{functionsmagneticBH} (using the supersymmetry conditions \eqref{magneticBHsusy}) and only depend on $\theta$,  while the horizon radii $r_{\pm}$ can be found in \eqref{horizonradii}.  Still using \eqref{toNH} and taking $\lambda \to 0$,  one can easily compute the gauge fields and the scalars on the horizon:
\begin{align}\label{AdS2fieldsfromBH}
\begin{split}
A_1&\, = \, \frac{(1-\alpha^2+\xp)(1-\alpha^2-\xp^2)}{\alpha^2\xp(1-\alpha^2+\xp\sqrt{1-\alpha^2}\cos\theta)}\diff \phi\,,\\
A_2&\, = \, \frac{(1-\alpha^2-\xp)(1-\alpha^2-\xp^2)}{\alpha^2\xp(1-\alpha^2-\xp\sqrt{1-\alpha^2}\cos\theta)}\diff \phi\,,\\
\ex^{\xi}&\, = \, \frac{(\alpha\sqrt{1-\alpha^2} r_+-\xp)(\sqrt{1-\alpha^2}+\xp \cos\theta)}{(\alpha \sqrt{1-\alpha^2} r_++\xp)(\sqrt{1-\alpha^2}-\xp \cos\theta)}\,, \qquad \chi\, = \, 0\,.
\end{split}
\end{align}
What is left to be shown is that the AdS$_2$ solution given by \eqref{AdS2metricfromBH} and \eqref{AdS2fieldsfromBH} is equivalent to \eqref{spinningspindle},  with $c_3=0=\jp$.  To this end,  we only need to find the change of coordinates from $y$ and $z$ in the latter to $\theta$ and $\phi$,  while also finding the map between the parameters $c_{1,2}$ in \eqref{spinningspindle} and $\alpha$,  $\xp$ used here.  For the coordinates,  we find
\begin{align}
y\, = \, \frac{t_1+t_2 \cos\theta}{1-\alpha r_+ \cos\theta}\,, \qquad z\, = \, - \kappa \phi\,,
\end{align}
with 
\begin{align}
\begin{split}
t_1&\, = \, -2\frac{r_+}{\sqrt{1-\alpha^2}(r_+-r_-)}\,,  \quad
t_2\, = \, \sqrt{1-\alpha^2}(r_+-r_-)\alpha c_2\,,\\
\kappa&\, = \, \frac{\sqrt{\xp^4-2 \xp^2 (1-\alpha^2)+(1-\alpha^2)(5-4 \alpha^2)}}{\alpha^2 (1-\alpha^2)}\,,
\end{split}
\end{align}
while the parameters are identified using
\begin{align}
\begin{split}
c_1&\, = \, -\frac{4(1-\alpha^2)^{3/2}(1-\alpha^2+\xp^2)}{\xp^4-2\xp^2\,(1-\alpha^2)+(1-\alpha^2)(5-4\alpha^2)}\,,\\
c_2&\, = \, \frac{2(1-\alpha^2)\xp^2}{\xp^4-2\xp^2(1-\alpha^2)+(1-\alpha^2)\,(5-4\alpha^2)}\,.
\end{split}
\end{align}
Finally,  we note that with the identifications above the gauge fields are equal only up to a gauge transformation.  In particular,  if we denote with $A_i^{AdS_2}$ the gauge fields in \eqref{spinningspindle} (with $c_3=0=\jp$) and with $A_i^{BH}$ the gauge fields in \eqref{AdS2fieldsfromBH},  we find
\begin{align}
A_i^{AdS_2}\, = \, A_i^{BH}-f_i\,\diff \phi\,,
\end{align}
with
\begin{align}
f_1\, = \, \frac{\xp(1-\alpha^2-\xp^2)+2(1-\alpha^2)^2}{2\xp\alpha^2(1-\alpha^2)}\,,\qquad
f_2\,  =\, \frac{\xp(1-\alpha^2-\xp^2)-2(1-\alpha^2)^2}{2\xp\alpha^2(1-\alpha^2)}\,.
\end{align}

\section{Accelerating black holes with 4  magnetic charges}\label{app:fourcharges}

An obvious extension of what was discussed in the body of this paper is to consider more general black holes,  with additional charges.  An interesting model where to look for such black holes is the STU model,  which can be seen alternatively as a $D=4$,  $\mathcal{N}=2$ supergravity coupled to three vector multiplets,  with prepotential $F=-\ii\,\sqrt{X^0\,X^1\,X^2\,X^3}$, or as a truncation of the maximal $D=4$,  $\mathcal{N}=8$ gauged supergravity to its abelian sector.  As such,  it can also be uplifted to $D=11$ supergravity,  making it interesting from an holographic viewpoint. The model discussed in section \ref{sec:model} can in fact be seen as a truncation of the STU model,  where the gauge fields are set to be pairwise equal,  while two of the scalars vanish.

Interestingly,  both an accelerating black hole solution and a local AdS$_2\times \Sigma$ solution can be found in the STU model,  with four distinct magnetic charges but no rotation or electric charge.  These two solutions will be the subject of this appendix and they can be extracted,  more or less directly,  from other existing solutions in the literature,  as we shall discuss momentarily.  In principle one could then perform the analysis that is found in the body of the present paper,  quantising the conical deficits of $\Sigma$ such that the horizon of the black hole is a spindle,  and computing the associated entropy.  However,  as we shall comment,  this is not easily done in practice,  due to the algebraic complication of the equations that one has to solve.

One feature of the solutions that we shall discuss is that,  since they are non rotating and only magnetically charged,  the axions of the STU model are set to zero.  This significantly simplifies the action of the theory,  therefore we shall only write the action with the three axions $\chi_i=0$.  This is given by
\begin{align}\label{actionU(1)^4}
\begin{split}
\mathcal{S}_{\mathrm{STU}\, ,\, \chi_i=0}\, =\, \int & (R-g^2\,\mathcal{V}) \star 1-\frac{1}{2}\sum_{i=1}^3 \diff \xi_i \wedge \star\xi_i-\frac{1}{2}\ex^{-\xi_1-\xi_2-\xi_3} F_1\wedge\star F_1\\
&-\frac{1}{2}\ex^{-\xi_1+\xi_2+\xi_3} F_2\wedge\star F_2-\frac{1}{2}\ex^{\xi_1-\xi_2+\xi_3} F_3\wedge\star F_3-\frac{1}{2}\ex^{\xi_1+\xi_2-\xi_3} F_4\wedge\star F_4\,,
\end{split}
\end{align}
where the scalar potential is
\begin{align}
\mathcal{V}\, = \, -2\,\left(\cosh \xi_1+\cosh\xi_2+\cosh\xi_3\right)\,,
\end{align}
and we shall from now on set the gauge coupling $g=1$,  as in the body of the paper.  Note that the action \eqref{actionU(1)^2} can be obtained from \eqref{actionU(1)^4} for example by setting
\begin{align}\label{4to2}
F_{1,2}\to F_2\,, \quad
F_{3,4}\to F_1\,, \quad
\xi_1\to\xi\,, \quad
\xi_{2,3}\to 0\,.
\end{align}

The supersymmetry variations of the fermionic fields are also simpler when the axions are set to zero.  We now have two additional vector multiplets compared to the discussion of section \ref{sec:model},  so the fermions of the theory are one gravitino $\psi_{\mu}$ and three gaugini $\lambda_i$ ($i=1,2,3$),  that we shall all take to be Dirac fermions.  From their supersymmetry variations,  we find the Killing spinor equations:
\begin{align}\label{KSE_STU}
\begin{split}
\delta\psi_{\mu}=&\left[\nabla_{\mu}-\frac{\ii}{4}\sum_{i=1}^4A_i+\frac{1}{8}\left(\frac{\ex^{\xi_1/2}}{\ex^{\xi_2/2}\,\ex^{\xi_3/2}}+\frac{\ex^{\xi_2/2}}{\ex^{\xi_1/2}\,\ex^{\xi_3/2}}+\frac{\ex^{\xi_3/2}}{\ex^{\xi_1/2}\,\ex^{\xi_2/2}}+\ex^{\xi_1/2}\,\ex^{\xi_2/2}\,\ex^{\xi_3/2}\right)\,\gamma_{\mu}\right.\\
&\left. +\frac{\ii}{16}\left(\frac{\slashed{F}_1}{\ex^{\xi_1/2}\,\ex^{\xi_2/2}\,\ex^{\xi_3/2}}+\frac{\ex^{\xi_2/2}\,\ex^{\xi_3/2}}{\ex^{\xi_1/2}}\,\slashed{F}_2+\frac{\ex^{\xi_1/2}\,\ex^{\xi_3/2}}{\ex^{\xi_2/2}}\,\slashed{F}_3+\frac{\ex^{\xi_1/2}\,\ex^{\xi_2/2}}{\ex^{\xi_3/2}}\,\slashed{F}_4\right)\,\gamma_{\mu}\right]\,\epsilon=0\,,\\
\delta \lambda_1=&\left[\ii\,\slashed{\partial}\xi_1-\frac{\ii}{2}\left(\frac{\ex^{\xi_1/2}}{\ex^{\xi_2/2}\,\ex^{\xi_3/2}}-\frac{\ex^{\xi_2/2}}{\ex^{\xi_1/2}\,\ex^{\xi_3/2}}-\frac{\ex^{\xi_3/2}}{\ex^{\xi_2/2}\,\ex^{\xi_2/2}}+\ex^{\xi_1/2}\,\ex^{\xi_2/2}\,\ex^{\xi_3/2}\right)\right.\\
&\left. +\frac{1}{4}\left(\frac{1}{\ex^{\xi_1/2}\,\ex^{\xi_2/2}\,\ex^{\xi_3/2}}\,\slashed{F}_1+\frac{\ex^{\xi_2/2}\,\ex^{\xi_3/2}}{\ex^{\xi_1/2}}\,\slashed{F}_2-\frac{\ex^{\xi_1/2}\,\ex^{\xi_3/2}}{\ex^{\xi_2/2}}\,\slashed{F}_3-\frac{\ex^{\xi_1/2}\,\ex^{\xi_2/2}}{\ex^{\xi_3/2}}\,\slashed{F}_4\right) \right]\,\epsilon=0\,,\\
\delta \lambda_2=&\left[\ii\,\slashed{\partial}\xi_2-\frac{\ii}{2}\left(-\frac{\ex^{\xi_1/2}}{\ex^{\xi_2/2}\,\ex^{\xi_3/2}}+\frac{\ex^{\xi_2/2}}{\ex^{\xi_1/2}\,\ex^{\xi_3/2}}-\frac{\ex^{\xi_3/2}}{\ex^{\xi_2/2}\,\ex^{\xi_2/2}}+\ex^{\xi_1/2}\,\ex^{\xi_2/2}\,\ex^{\xi_3/2}\right)\right.\\
&\left. +\frac{1}{4}\left(\frac{1}{\ex^{\xi_1/2}\,\ex^{\xi_2/2}\,\ex^{\xi_3/2}}\,\slashed{F}_1-\frac{\ex^{\xi_2/2}\,\ex^{\xi_3/2}}{\ex^{\xi_1/2}}\,\slashed{F}_2+\frac{\ex^{\xi_1/2}\,\ex^{\xi_3/2}}{\ex^{\xi_2/2}}\,\slashed{F}_3-\frac{\ex^{\xi_1/2}\,\ex^{\xi_2/2}}{\ex^{\xi_3/2}}\,\slashed{F}_4\right) \right]\,\epsilon=0\,,\\
\delta \lambda_3=&\left[\ii\,\slashed{\partial}\xi_3-\frac{\ii}{2}\left(-\frac{\ex^{\xi_1/2}}{\ex^{\xi_2/2}\,\ex^{\xi_3/2}}-\frac{\ex^{\xi_2/2}}{\ex^{\xi_1/2}\,\ex^{\xi_3/2}}+\frac{\ex^{\xi_3/2}}{\ex^{\xi_2/2}\,\ex^{\xi_2/2}}+\ex^{\xi_1/2}\,\ex^{\xi_2/2}\,\ex^{\xi_3/2}\right)\right.\\
&\left. +\frac{1}{4}\left(\frac{1}{\ex^{\xi_1/2}\,\ex^{\xi_2/2}\,\ex^{\xi_3/2}}\,\slashed{F}_1-\frac{\ex^{\xi_2/2}\,\ex^{\xi_3/2}}{\ex^{\xi_1/2}}\,\slashed{F}_2-\frac{\ex^{\xi_1/2}\,\ex^{\xi_3/2}}{\ex^{\xi_2/2}}\,\slashed{F}_3+\frac{\ex^{\xi_1/2}\,\ex^{\xi_2/2}}{\ex^{\xi_3/2}}\,\slashed{F}_4\right) \right]\,\epsilon=0\,,
\end{split}
\end{align}
which must all be satisfied simultaneously by a Dirac fermion $\epsilon$ for a bosonic solution to preserve some supersymmetry.  Note that with the replacements \eqref{4to2} these equations reduce to the Killing spinor equations for the theory with two gauge fields given in \eqref{KSEs},  in the case of vanishing axions.

\subsection{Accelerating magnetic black holes}

We remind that in section \ref{sec:BH} we started from accelerating charged black hole solutions with two electric presented in \cite{Lu:2014sza},  and constructed analogous black holes with two magnetic charges using electromagnetic duality.  Likewise,  one could dualize the accelerating black holes with four electric charges given in \cite{Lu:2014sza},  which are solutions of the STU model with vanishing axions,  to obtain accelerating black holes with four magnetic charges.  We write the solution as 
\begin{align}\label{4chargeBH}
\begin{split}
\diff s^2\,=\, &\frac{1}{H^2}\left[-\frac{f^{1/2}}{h^{1/2}}\,Y\,\diff t^2+f^{1/2}\,h^{1/2}\,\frac{\diff y^2}{Y}+f^{1/2}\,h^{1/2}\,\frac{\diff x^2}{X}+\frac{h^{1/2}}{f^{1/2}}\,X\,\diff \phi^2\right]\,,\\
A_i=&\frac{\sqrt{p_i^2-\mu}}{\alpha\,p_i\,f_i}\,\diff \phi\,,\\
\ex^{\xi_1}\, = \, &\left(\frac{h_1\,h_2\,f_3\,f_4}{h_3\,h_4\,f_1\,f_2}\right)^{1/2}\,,  \quad
\ex^{\xi_2}\, = \, \left(\frac{h_1\,h_3\,f_2\,f_4}{h_2\,h_4\,f_1\,f_3}\right)^{1/2}\,, \quad
\ex^{\xi_3}\, = \, \left(\frac{h_1\,h_4\,f_2\,f_3}{h_2\,h_3\,f_1\,f_4}\right)^{1/2}\,, 
\end{split}
\end{align}
where we have introduced the functions
\begin{align}
\begin{split}
Y\, = \, &\,y^2-\mu\,y^3+(1-\alpha^2)\,h\quad
X\, = \, -x^2\,(1-\alpha\,\mu\,x)+f\,,\quad
H=y-\alpha\,x\,,\\
h\, = \, &\prod_{i=1}^4 h_i\,, \qquad f \, = \, \prod_{i=1}^4 f_i\,, \qquad
h_i\, = \, 1-p_i^2\,y\,, \qquad
f_i\, = \, 1-\alpha\,p_i^2\,x\,.
\end{split}
\end{align}
In the above,  one should think of $y$ as the inverse of the radial coordinate of a black hole (so that the curvature singularity is located at $y=\infty$),  and of $x$ as an angular variable (one should set $x=a\,\cos\theta+b$ for some $a$ and $b$).  These coordinates are particularly convenient as they simplify the study of the equations of motion and of the supersymmetry conditions.  As written above,  the solution is not supersymmetric and is given in terms of six parameters: $\mu$,  related to the mass of the black hole,  $\alpha$,  related to the acceleration,  and $p_i$ ($i=1,\dots 4$),  related to the magnetic charges.

After some technical manipulations,  one can show that the vanishing of the gaugino variations in \eqref{KSE_STU} imply that a necessary condition for supersymmetry is that the parameters are constrained by
\begin{align}\label{BPS4chargeBH}
\begin{split}
\alpha&\,=\,\left(1+\frac{\mu}{(p_1\,p_2+p_3\,p_4)\,(p_1\,p_3+p_2\,p_4)\,(p_1\,p_4+p_2\,p_3)}\right)^{1/2}\,, \\
\mu&\,=\,\frac{4\,(p_1\,p_2+p_3\,p_4)\,(p_1\,p_3+p_2\,p_4)\,(p_1\,p_4+p_2\,p_3)}{(p_1+p_2+p_3-p_4)\,(p_1+p_2-p_3+p_4)\,(p_1-p_2+p_3+p_4)\,(-p_1+p_2+p_3+p_4)}\,,
\end{split}
\end{align}
which we also expect to be sufficient for supersymmetry to be preserved.  The resulting supersymmetric black hole is also extremal,  and is given in terms of four independent magnetic charge parameters $p_i$,  $i=1,\dots 4$.  Its near-horizon limit is then a supersymmetric AdS$_2\times \Sigma$ solution,  which we conjecture to be the one that we discuss in the next subsection.  One could in principle prove the equivalence of the two with a direct computation,   but this is algebraically involved and we leave this check for future work.

\subsection{AdS$_2$ spindle with four magnetic charges}

Let us now present a class of supersymmetric AdS$_2$ solutions of the STU model,  with four magnetic charges, that we conjecture to arise as the near horizon limit of the black holes \eqref{4chargeBH},  in the supersymmetric and extremal limit given by the BPS conditions \eqref{BPS4chargeBH}.  Such solutions can be extracted from the supersymmetric solutions of $D=11$ supergravity given in section 5.4 of \cite{Gauntlett:2006ns},  and they reduce to the two-charge AdS$_2$ solutions \eqref{2chargemagneticAdS2} in the case of pairwise equal charges.  Using the uplifting formulas of \cite{Azizi:2016noi},  we find
\begin{align}\label{4chargemagneticAdS2}
\begin{split}
\diff s^2_4&\, = \, \frac{1}{4}\Lambda(w)\left(-\rho^2\diff\tau^2+\frac{\diff\rho^2}{\rho^2}\right)+\frac{\Lambda(w)}{Q(w)}\diff w^2+\frac{Q(w)}{4\Lambda(w)}\diff \z^2\,,\\
A_i&\, =\, \frac{w}{w+q_i}\diff z\,, \hspace{2.8cm}
\ex^{\xi_1}\, = \, \left[\frac{(w+q_3)(w+q_4)}{(w+q_1)(w+q_2)}\right]^{1/2}\,,\\
\ex^{\xi_2}&\, = \, \left[\frac{(w+q_2)(w+q_4)}{(w+q_1)(w+q_3)}\right]^{1/2}\,,\quad
\ex^{\xi_3}\, = \, \left[\frac{(w+q_2)(w+q_3)}{(w+q_1)(w+q_4)}\right]^{1/2}\,,
\end{split}
\end{align}
where
\begin{align}\label{functions4chargeAdS2}
\Lambda(w)\, = \, \left[(w+q_1)(w+q_2)(w+q_3)(w+q_4)\right]^{1/2}\,, \qquad
Q(w)\, = \, \Lambda(w)^2-4w^2\,.
\end{align}
We have checked explicitly that this solution is indeed supersymmetric,  and its two linearly independent Killing spinors $\epsilon_{1,2}$ can be expressed as in \eqref{killingspinors} in terms of two two-dimensional spinors $\eta_{1,2}$,  which are given by
\begin{align}
\eta_1\, = \, \frac{\ex^{\tfrac{\ii \,z}{2}}}{\Lambda(w)^{1/4}}\begin{pmatrix}
\sqrt{\Lambda(w)-2\,w}\\
\sqrt{\Lambda(w)+2\,w}
\end{pmatrix}
\,, \qquad
\eta_2\, = \, \ii\,\frac{\ex^{\tfrac{\ii \,z}{2}}}{\Lambda(w)^{1/4}}\begin{pmatrix}
-\sqrt{\Lambda(w)-2\,w}\\
\sqrt{\Lambda(w)+2\,w}
\end{pmatrix}\,.
\end{align}
While we have not attacked it,  an interesting open problem is that of generalizing the solution (\ref{4chargemagneticAdS2}) -- (\ref{functions4chargeAdS2}) to include electric charges and rotation,  in the spirit of what we have done in section \ref{sec:AdS2solutions}.

\providecommand{\href}[2]{#2}\begingroup\raggedright\endgroup


\begin{thebibliography}{10}

\bibitem{Benini:2015eyy}
F.~Benini, K.~Hristov, and A.~Zaffaroni, {\it {Black hole microstates in
  AdS$_{4}$ from supersymmetric localization}},  {\em JHEP} {\bf 05} (2016)
  054, [\href{http://arxiv.org/abs/1511.04085}{{\tt arXiv:1511.04085}}].

\bibitem{Benini:2016rke}
F.~Benini, K.~Hristov, and A.~Zaffaroni, {\it {Exact microstate counting for
  dyonic black holes in AdS4}},  {\em Phys. Lett. B} {\bf 771} (2017) 462--466,
  [\href{http://arxiv.org/abs/1608.07294}{{\tt arXiv:1608.07294}}].

\bibitem{Benini:2015noa}
F.~Benini and A.~Zaffaroni, {\it {A topologically twisted index for
  three-dimensional supersymmetric theories}},  {\em JHEP} {\bf 07} (2015) 127,
  [\href{http://arxiv.org/abs/1504.03698}{{\tt arXiv:1504.03698}}].

\bibitem{Nian:2019pxj}
J.~Nian and L.~A. Pando~Zayas, {\it {Microscopic entropy of rotating
  electrically charged AdS$_{4}$ black holes from field theory localization}},
  {\em JHEP} {\bf 03} (2020) 081, [\href{http://arxiv.org/abs/1909.07943}{{\tt
  arXiv:1909.07943}}].

\bibitem{Ferrero:2020twa}
P.~Ferrero, J.~P. Gauntlett, J.~M.~P. Ipi\~na, D.~Martelli, and J.~Sparks, {\it
  {Accelerating black holes and spinning spindles}},  {\em Phys. Rev. D} {\bf
  104} (2021), no.~4 046007, [\href{http://arxiv.org/abs/2012.08530}{{\tt
  arXiv:2012.08530}}].

\bibitem{Plebanski:1976gy}
J.~F. Plebanski and M.~Demianski, {\it {Rotating, charged, and uniformly
  accelerating mass in general relativity}},  {\em Annals Phys.} {\bf 98}
  (1976) 98--127.

\bibitem{Podolsky:2006px}
J.~Podolsky and J.~B. Griffiths, {\it {Accelerating Kerr-Newman black holes in
  (anti-)de Sitter space-time}},  {\em Phys. Rev. D} {\bf 73} (2006) 044018,
  [\href{http://arxiv.org/abs/gr-qc/0601130}{{\tt gr-qc/0601130}}].

\bibitem{PhysRevD.2.1359}
W.~Kinnersley and M.~Walker, {\it Uniformly accelerating charged mass in
  general relativity},  {\em Phys. Rev. D} {\bf 2} (Oct, 1970) 1359--1370.

\bibitem{Podolsky:2003gm}
J.~Podolsky, M.~Ortaggio, and P.~Krtous, {\it {Radiation from accelerated black
  holes in an anti-de Sitter universe}},  {\em Phys. Rev. D} {\bf 68} (2003)
  124004, [\href{http://arxiv.org/abs/gr-qc/0307108}{{\tt gr-qc/0307108}}].

\bibitem{Klemm:2013eca}
D.~Klemm and M.~Nozawa, {\it {Supersymmetry of the C-metric and the general
  Plebanski-Demianski solution}},  {\em JHEP} {\bf 05} (2013) 123,
  [\href{http://arxiv.org/abs/1303.3119}{{\tt arXiv:1303.3119}}].

\bibitem{Bobev:2019zmz}
N.~Bobev and P.~M. Crichigno, {\it {Universal spinning black holes and theories
  of class $ \mathcal{R} $}},  {\em JHEP} {\bf 12} (2019) 054,
  [\href{http://arxiv.org/abs/1909.05873}{{\tt arXiv:1909.05873}}].

\bibitem{Gauntlett:2006ns}
J.~P. Gauntlett, N.~Kim, and D.~Waldram, {\it {Supersymmetric AdS(3), AdS(2)
  and Bubble Solutions}},  {\em JHEP} {\bf 04} (2007) 005,
  [\href{http://arxiv.org/abs/hep-th/0612253}{{\tt hep-th/0612253}}].

\bibitem{Ferrero:2020laf}
P.~Ferrero, J.~P. Gauntlett, J.~M. P\'erez Ipi\~na, D.~Martelli, and J.~Sparks,
  {\it {D3-Branes Wrapped on a Spindle}},  {\em Phys. Rev. Lett.} {\bf 126}
  (2021), no.~11 111601, [\href{http://arxiv.org/abs/2011.10579}{{\tt
  arXiv:2011.10579}}].

\bibitem{Cassani:2021dwa}
D.~Cassani, J.~P. Gauntlett, D.~Martelli, and J.~Sparks, {\it {Thermodynamics
  of accelerating and supersymmetric AdS4 black holes}},  {\em Phys. Rev. D}
  {\bf 104} (2021), no.~8 086005, [\href{http://arxiv.org/abs/2106.05571}{{\tt
  arXiv:2106.05571}}].

\bibitem{Cabo-Bizet:2018ehj}
A.~Cabo-Bizet, D.~Cassani, D.~Martelli, and S.~Murthy, {\it {Microscopic origin
  of the Bekenstein-Hawking entropy of supersymmetric AdS$_{5}$ black holes}},
  {\em JHEP} {\bf 10} (2019) 062, [\href{http://arxiv.org/abs/1810.11442}{{\tt
  arXiv:1810.11442}}].

\bibitem{Cassani:2019mms}
D.~Cassani and L.~Papini, {\it {The BPS limit of rotating AdS black hole
  thermodynamics}},  {\em JHEP} {\bf 09} (2019) 079,
  [\href{http://arxiv.org/abs/1906.10148}{{\tt arXiv:1906.10148}}].

\bibitem{Gaiotto:2008ak}
D.~Gaiotto and E.~Witten, {\it {S-Duality of Boundary Conditions In N=4 Super
  Yang-Mills Theory}},  {\em Adv. Theor. Math. Phys.} {\bf 13} (2009), no.~3
  721--896, [\href{http://arxiv.org/abs/0807.3720}{{\tt arXiv:0807.3720}}].

\bibitem{Lu:2014sza}
H.~L\"u and J.~F. V\'azquez-Poritz, {\it {C-metrics in Gauged STU Supergravity
  and Beyond}},  {\em JHEP} {\bf 12} (2014) 057,
  [\href{http://arxiv.org/abs/1408.6531}{{\tt arXiv:1408.6531}}].

\bibitem{Chow:2013gba}
D.~D.~K. Chow and G.~Comp\`ere, {\it {Dyonic AdS black holes in maximal gauged
  supergravity}},  {\em Phys. Rev. D} {\bf 89} (2014), no.~6 065003,
  [\href{http://arxiv.org/abs/1311.1204}{{\tt arXiv:1311.1204}}].

\bibitem{Hristov:2019mqp}
K.~Hristov, S.~Katmadas, and C.~Toldo, {\it {Matter-coupled supersymmetric
  Kerr-Newman-AdS$_4$ black holes}},  {\em Phys. Rev. D} {\bf 100} (2019),
  no.~6 066016, [\href{http://arxiv.org/abs/1907.05192}{{\tt
  arXiv:1907.05192}}].

\bibitem{Chong:2004na}
Z.~W. Chong, M.~Cvetic, H.~Lu, and C.~N. Pope, {\it {Charged rotating black
  holes in four-dimensional gauged and ungauged supergravities}},  {\em Nucl.
  Phys. B} {\bf 717} (2005) 246--271,
  [\href{http://arxiv.org/abs/hep-th/0411045}{{\tt hep-th/0411045}}].

\bibitem{Cvetic:2005zi}
M.~Cvetic, G.~W. Gibbons, H.~Lu, and C.~N. Pope, {\it {Rotating black holes in
  gauged supergravities: Thermodynamics, supersymmetric limits, topological
  solitons and time machines}},
  \href{http://arxiv.org/abs/hep-th/0504080}{{\tt hep-th/0504080}}.

\bibitem{Caldarelli:1998hg}
M.~M. Caldarelli and D.~Klemm, {\it {Supersymmetry of Anti-de Sitter black
  holes}},  {\em Nucl. Phys. B} {\bf 545} (1999) 434--460,
  [\href{http://arxiv.org/abs/hep-th/9808097}{{\tt hep-th/9808097}}].

\bibitem{Faedo:2021kur}
F.~Faedo, S.~Klemm, and A.~Vigan\`o, {\it {Supersymmetric black holes with
  spiky horizons}},  {\em JHEP} {\bf 09} (2021) 102,
  [\href{http://arxiv.org/abs/2105.02902}{{\tt arXiv:2105.02902}}].

\bibitem{Hosseini:2020mut}
S.~M. Hosseini and A.~Zaffaroni, {\it {Universal AdS Black Holes in Theories
  with 16 Supercharges and Their Microstates}},  {\em Phys. Rev. Lett.} {\bf
  126} (2021), no.~17 171604, [\href{http://arxiv.org/abs/2011.01249}{{\tt
  arXiv:2011.01249}}].

\bibitem{deWit:1984wbb}
B.~de~Wit and A.~Van~Proeyen, {\it {Potentials and Symmetries of General Gauged
  N=2 Supergravity: Yang-Mills Models}},  {\em Nucl. Phys. B} {\bf 245} (1984)
  89--117.

\bibitem{Cvetic:1999xp}
M.~Cvetic, M.~J. Duff, P.~Hoxha, J.~T. Liu, H.~Lu, J.~X. Lu,
  R.~Martinez-Acosta, C.~N. Pope, H.~Sati, and T.~A. Tran, {\it {Embedding AdS
  black holes in ten-dimensions and eleven-dimensions}},  {\em Nucl. Phys. B}
  {\bf 558} (1999) 96--126, [\href{http://arxiv.org/abs/hep-th/9903214}{{\tt
  hep-th/9903214}}].

\bibitem{Cacciatori:2008ek}
S.~L. Cacciatori, D.~Klemm, D.~S. Mansi, and E.~Zorzan, {\it {All timelike
  supersymmetric solutions of N=2, D=4 gauged supergravity coupled to abelian
  vector multiplets}},  {\em JHEP} {\bf 05} (2008) 097,
  [\href{http://arxiv.org/abs/0804.0009}{{\tt arXiv:0804.0009}}].

\bibitem{Azizi:2016noi}
A.~Azizi, H.~Godazgar, M.~Godazgar, and C.~N. Pope, {\it {Embedding of gauged
  STU supergravity in eleven dimensions}},  {\em Phys. Rev. D} {\bf 94} (2016),
  no.~6 066003, [\href{http://arxiv.org/abs/1606.06954}{{\tt
  arXiv:1606.06954}}].

\bibitem{Fujii:1985bg}
Y.~Fujii and K.~Yamagishi, {\it {Killing spinors on spheres and hyperbolic
  manifolds}},  {\em J. Math. Phys.} {\bf 27} (1986) 979.

\bibitem{Ferrero:2021wvk}
P.~Ferrero, J.~P. Gauntlett, D.~Martelli, and J.~Sparks, {\it {M5-branes
  wrapped on a spindle}},  {\em JHEP} {\bf 11} (2021) 002,
  [\href{http://arxiv.org/abs/2105.13344}{{\tt arXiv:2105.13344}}].

\bibitem{Ferrero:2021etw}
P.~Ferrero, J.~P. Gauntlett, and J.~Sparks, {\it {Supersymmetric spindles}},
  {\em JHEP} {\bf 01} (2022) 102, [\href{http://arxiv.org/abs/2112.01543}{{\tt
  arXiv:2112.01543}}].

\bibitem{BenettiGenolini:2019jdz}
P.~Benetti~Genolini, J.~M. Perez Ipi\~na, and J.~Sparks, {\it {Localization of
  the action in AdS/CFT}},  {\em JHEP} {\bf 10} (2019) 252,
  [\href{http://arxiv.org/abs/1906.11249}{{\tt arXiv:1906.11249}}].

\bibitem{Anabalon:2018qfv}
A.~Anabal\'on, F.~Gray, R.~Gregory, D.~Kubizn\'ak, and R.~B. Mann, {\it
  {Thermodynamics of Charged, Rotating, and Accelerating Black Holes}},  {\em
  JHEP} {\bf 04} (2019) 096, [\href{http://arxiv.org/abs/1811.04936}{{\tt
  arXiv:1811.04936}}].

\bibitem{Couzens:2020jgx}
C.~Couzens, E.~Marcus, K.~Stemerdink, and D.~van~de Heisteeg, {\it {The
  near-horizon geometry of supersymmetric rotating AdS$_{4}$ black holes in
  M-theory}},  {\em JHEP} {\bf 05} (2021) 194,
  [\href{http://arxiv.org/abs/2011.07071}{{\tt arXiv:2011.07071}}].

\bibitem{Hosseini:2021fge}
S.~M. Hosseini, K.~Hristov, and A.~Zaffaroni, {\it {Rotating multi-charge
  spindles and their microstates}},  {\em JHEP} {\bf 07} (2021) 182,
  [\href{http://arxiv.org/abs/2104.11249}{{\tt arXiv:2104.11249}}].

\bibitem{Boido:2021szx}
A.~Boido, J.~M.~P. Ipi\~na, and J.~Sparks, {\it {Twisted D3-brane and M5-brane
  compactifications from multi-charge spindles}},  {\em JHEP} {\bf 07} (2021)
  222, [\href{http://arxiv.org/abs/2104.13287}{{\tt arXiv:2104.13287}}].

\end{thebibliography}
\end{document}